\documentclass[12pt,a4paper]{article}

\usepackage{amsmath,amssymb,slashed,cancel}
\usepackage{cite,tabularx}
\usepackage{subcaption}
\usepackage{graphicx,color}
\usepackage[normalem]{ulem}
\usepackage{mathtools}
\usepackage{enumerate}
\usepackage{ascmac}
\allowdisplaybreaks

\usepackage[height=8.85in,width=6.4in]{geometry}
\renewcommand{\baselinestretch}{1.1}
\newcommand\package[2][\relax]{\texttt{#2\ifx#1\relax\relax\relax\else\,\linebreak[0]#1\fi}}

\numberwithin{equation}{section} 
 
\def\beq#1\eeq{\begin{align}#1\end{align}}

\definecolor{BlueViolet}{rgb}{0.2, 0.00, 0.7}
\definecolor{Blue}{rgb}{0.15, 0.00, 0.9}
\usepackage[colorlinks=true,linkcolor=Blue,citecolor=Blue,urlcolor=BlueViolet]{hyperref}

\usepackage{comment}

\begin{document}
\begin{titlepage}
\setcounter{page}{0} 

\begin{center}

\vskip .55in

\begingroup
\centering
\large\bf Quantum Resonance Beyond Direct Measurement:
\\
Insights from Weak Measurement
\endgroup

\vskip .4in

{
  Daiki Ueda$^{\rm (a)}$ and Izumi Tsutsui$^{\rm (b,c)}$
}

\vskip 0.4in

\begingroup\small
\begin{minipage}[t]{0.9\textwidth}
\centering\renewcommand{\arraystretch}{0.9}
\begin{tabular}{c@{\,}l}
$^{\rm(a)}$
& Physics Department, Technion \text{--} Israel Institute of Technology,
Technion city,\\
&Haifa 3200003, Israel
\\ [2mm]
$^{\rm(b)}$
& Department of Physics, College of Science and Technology, Nihon University,
\\
&Tokyo 101-8308, Japan\\
$^{\rm (c)}$
& Theory Center, IPNS, High Energy Accelerator Research Organization (KEK),
\\
& Tsukuba 305-0801, Japan
\end{tabular}
\end{minipage}
\endgroup

\end{center}

\vskip .4in

\begin{abstract}\noindent
Aharonov's weak value amplification realized via weak measurement provides a versatile means to measure physical parameters with high accuracy, similar to   
quantum resonance epitomized by the Rabi and Ramsey resonances. The similarity between the two is not accidental: in fact, these two methods of precision measurement have recently been shown to be interconnected for the particular case of direct weak measurement. Here we show that this connection also holds in the case of indirect weak measurement, which is the standard scheme used mostly for studies of weak value amplification. We present a unified framework of weak measurements in which direct measurement appears as a special case of indirect measurement. This allows us to compare the measurement precision of different methods on a common basis, as demonstrated explicitly for the Rabi and Ramsey resonances, showing how the precision of the latter surpasses that of the former in the context of weak value amplification.
\end{abstract}
\end{titlepage}

\setcounter{page}{1}
\renewcommand{\thefootnote}{\#\arabic{footnote}}
\setcounter{footnote}{0}

\begingroup
\renewcommand{\baselinestretch}{1} 
\setlength{\parskip}{2pt}          
\hrule
\tableofcontents
\vskip .2in
\hrule
\vskip .4in
\endgroup

\newpage
\section{Introduction}	
Progress in physics requires both the discovery of fundamental constituents and the precise determination of their properties.
Resonance phenomena contribute to both aims, serving as exceptionally sensitive probes of nature at its most fundamental level.
Resonance is a ubiquitous feature of oscillatory systems, emerging whenever specific conditions are satisfied, as exemplified by a pendulum under periodic driving or microwave cavities in electromagnetic systems.
This universality has led to a wide range of applications, ranging from the measurement of sound frequencies via acoustic resonance to the realization of filters in electrical circuits through electrical resonance, and also to the search of new unstable \lq resonance\rq\ particles formed in scattering experiment.    
In fundamental quantum physics, resonance plays an especially critical role in the accurate determination of physical parameters.
For example, in the International System of Units (SI)~\cite{sibrochure,247071}, the definition of the second is based on the ground-state hyperfine transition frequency of cesium-133 atoms, while in particle physics, quantities such as the neutron electric dipole moment (EDM)~\cite{Abel:2020pzs} are determined using resonance-based techniques.

Quantum resonance arises as an enhancement of the transition probability between initial and final states.
A prominent example is magnetic resonance, first explored by Rabi in 1937 within a general theoretical framework~\cite{Rabi:1937dgo} and subsequently applied to the precision measurement of nuclear magnetic moments~\cite{PhysRev.53.318,PhysRev.53.495,Rabi:1939jho}.
In 1950, Ramsey introduced a new resonance mechanism~\cite{Ramsey:1950tr}, later confirmed experimentally, which offers distinct advantages over the Rabi resonance.
Notably, the Ramsey resonance exhibits a narrower half-width (approximately 0.6 times that of the Rabi resonance), yielding greater sensitivity to deviations from the resonance condition.
Moreover, it is less susceptible to perturbations from external electric or magnetic fields, further enhancing its precision and robustness.

An alternative approach to precision measurement of fundamental parameters is provided by weak measurement which was introduced by Aharonov, Albert, and Vaidman in 1988~\cite{PhysRevLett.60.1351}.
In this framework, a physical quantity is measured weakly, with the condition that the state of the system be found in a prescribed post-selected state at the end of the process.
The resulting quantity, known as the weak value, offers novel insights on the physical properties of the system, which are naturally interpreted in the time-symmetric description of the measurement process involving post-selection~\cite{Aharonov2008}.
Since weak values enable the inference on the status of observables under minimal disturbance to the system, they have been used to shed light on various quantum paradoxes~\cite{aredes2024association}, including the Three Box Paradox~\cite{Aharonov1991CompleteDO}, Danan’s paradox concerning photon trajectories~\cite{PhysRevLett.111.240402, Vaidman-Tsutsui}, and the quantum Cheshire Cat phenomenon~\cite{Denkmayr2014,aharonov2021dynamical,aharonov2024angular,Ghoshal:2022bnc}.

From a practical viewpoint, optical implementations have demonstrated that weak measurement can be harnessed for weak value amplification~\cite{PhysRevLett.66.1107,Dressel_2014}.
This merit has been successfully utilized for precision measurements, such as the observation of the spin Hall effect of light~\cite{hosten2008observation} and the detection of ultrasensitive beam deflections in a Sagnac interferometer~\cite{PhysRevLett.102.173601}.
A prominent feature of weak value amplification is that it comes at the cost of suppressing the transition probability between the initial (pre-selected) and final (post-selected) states.
Interestingly, quantum resonance and weak value amplification share an analogous feature: in both cases, transition probabilities can be either enhanced or suppressed under suitable conditions.
Nonetheless, despite extensive investigations of these two amplification strategies in precision measurements, their interrelation has remained largely unexplored.
This motivates us to reconsider quantum resonance within the framework of weak measurement to elucidate possible connections between the two.

In this direction, in our previous work~\cite{Ueda:2023xoj} we revisited the Rabi and Ramsey resonances and demonstrated that they can be interpreted as instances of weak value amplification in direct measurements, where transition probabilities are significantly enhanced.
We also observed that, near their respective resonance points, the transition probabilities exhibit an almost identical behavior, differing only in sensitivity, which scales with the measurement strength.
In particular, the Ramsey resonance was found to correspond to a weak measurement up to a factor of $\pi/2\simeq 1/0.6$ stronger than the Rabi resonance, consistent with the half-width relation noted above.
Besides, we argued that previous neutron EDM experiments may have effectively determined the imaginary part of the weak value of the neutron spin with a precision exceeding that of standard weak value measurements with neutron beams by three orders of magnitude.

In general, weak measurements can be classified into two types: direct and indirect. The previous argument employed the direct scheme, which is simpler and offers a clearer picture of the correspondence between quantum resonance and weak value amplification.
However, given that the scope of application is rather limited for direct measurement as far as the amplification is concerned, a natural question arises as to whether a counterpart of quantum resonance exists in the indirect weak measurement, which is the standard scheme adopted in most of the precision measurements discussed above.
To address this question, in this paper we present a unified framework of weak measurement that encompasses both direct and indirect schemes. 
Within this framework, we identify counterparts of the Rabi and Ramsey resonances in the indirect weak measurement, which reduce to the conventional Rabi and Ramsey resonances as special cases.
This suggests that any resonance observed in direct measurement has a counterpart in indirect measurement, and vice versa. 
In this sense, weak measurement may open a pathway to discovering new quantum resonances, arising from extensions beyond the direct measurement scheme.

This paper is organized as follows.
Section~\ref{sec:weak} reviews indirect and direct weak measurements and then provides a unified framework of weak measurement that incorporates both direct and indirect schemes, clarifying their relation within a general formalism.
Section~\ref{sec:weak_qubit} presents a general study of weak measurement in two-state (qubit) systems, encompassing both direct and indirect schemes, with the aim of elucidating their correspondence with quantum resonance.
In this section, weak measurements are classified into  commutative and non-commutative types according to the form of the Hamiltonian governing the qubit system.
Section~\ref{sec:reso} presents a concise review of quantum resonance, focusing on two representative examples, the Rabi and Ramsey resonances, to facilitate understanding of their correspondence with weak measurement.
Section~\ref{sec:weak_reso} begins with a review of Ref.\cite{Ueda:2023xoj} aiming at clarifying the correspondence between conventional quantum resonance and direct weak measurement, and subsequently extends the discussion to indirect measurement to demonstrate the counterparts of the Rabi and Ramsey resonances in this context.
Finally, Section~\ref{sec:sum} is devoted to summary and discussions.

\section{Indirect and direct measurements}
\label{sec:weak}

The indirect weak measurement~\cite{PhysRevLett.60.1351,PhysRevA.41.11,PhysRevLett.62.2326,hosten2008observation,PhysRevLett.102.173601,PhysRevA.82.063822,lee2013uncertainty,Denkmayr2014} is arguably the standard scheme used when we discuss the weak value~\cite{PhysRev.134.B1410,PhysRevLett.60.1351} of a physical observable $A$ in the system ${\cal S}$ of our concern.  It presumes a setup of indirect measurement, in which we employ a probe ${\cal P}$ coupled to the system ${\cal S}$ through the conventional von Neumann measurement interaction Hamiltonian~\cite{vonNeumann1955},
\begin{align}
H = g A \otimes P,\label{eq:ham}
\end{align}
with an observable $P$ of the probe system ${\cal P}$.  
Here, the coupling constant $g$ is assumed to be sufficiently small so that the first order approximation is admitted in the following argument.  We shall begin our argument by recalling how the weak value of the observable $A$ arises in this basic setup.

Given the initial state ${|\psi_i\rangle }$ (called the pre-selected state) of the system ${\cal S}$ and the initial state ${|\Psi_i\rangle}$ of the probe ${\cal P}$,  during the time duration $t$ the combined system undergoes the evolution (we put $\hbar = 1$ for simplicity),
\begin{align}
{|\psi_i\otimes\Psi_i\rangle} \longrightarrow e^{-iH t}  {|\psi_i\otimes\Psi_i\rangle} = e^{-igt  A \otimes  P}  {|\psi_i\otimes\Psi_i\rangle}.\label{eq:tev}
\end{align}
The post-selection, {\it i.e.}, the procedure implementing the selection of the normalized state ${|\psi_f\rangle}$ (called the post-selected state) in the system ${\cal S}$ after the time duration then yields the final (non-normalized) state of the probe ${\cal P}$,
\begin{align}
{|\Psi_f\rangle} := {\langle \psi_f|} e^{-igt  A \otimes  P}  {|\psi_i\otimes\Psi_i\rangle}.\label{eq:pss}
\end{align}
Keeping only the leading term linear in $g$ in the expansion of the exponential, we find 
\begin{align}
{|\Psi_f\rangle} \simeq {\langle\psi_f|} I_{\mathcal{S}}\otimes I_{\mathcal{P}}  -igt A \otimes P  {|\psi_i\otimes\Psi_i\rangle} = {\langle \psi_f|\psi_i\rangle}\left(I_{\mathcal{P}}-igt A_w  P\right){|\Psi_i\rangle},\label{eq:psstwo}
\end{align}
where $I_{\mathcal{S}}$ and $I_{\mathcal{P}}$ are the identity operators on $\mathcal{S}$ and $\mathcal{P}$, respectively, and
\begin{align}
A_w = \frac{{\langle\psi_f|} A {|\psi_i\rangle}}{\langle\psi_f|\psi_i\rangle},\label{eq:wv}
\end{align}
is the quantity called the {\it weak value} of the observable $ A$.  Although this is the familiar procedure to retrieve the weak value for $A$, there are cases where the weak value may be retrieved without introducing the probe $\mathcal{P}$.  Before discussing this direct measurement, we clarify what we mean by indirect measurement in more detail.


\subsection{Indirect measurement}
\label{sec:Ind}
%
Following the conventional protocol of measuring physical observables, we wish to measure the weak value via the shift of the pointer of the probe.  Let $F$ be an observable of the probe which is deemed to represent the \lq position\rq\ of the pointer of the measuring device.  The measurement is then performed by comparing the position of the pointer observed at the time of the pre-selection and that observed at the time of the post-selection, that is, we measure $F$ for both the states ${|\Psi_i\rangle}$ and ${|\Psi_f\rangle}$ successively, and compare the difference in the expectation values between the two measurement outcomes.  

To see how the outcomes will turn out, we first use Eq.~\eqref{eq:psstwo} to find 
\begin{align}
{\langle \Psi_f|}  F {|\Psi_f\rangle} 
&\simeq  
\vert {\langle\psi_f| \psi_i\rangle} \vert^2 \notag
\\
&\times\bigg\{{\langle\Psi_i|}  F {|\Psi_i\rangle}  + gt\bigg(\,{\rm Re}\, A_w {\langle \Psi_i|} C_{FP} {|\Psi_i\rangle} 
+ \,{\rm Im}\, A_w {\langle\Psi_i|} A_{FP} {|\Psi_i\rangle}\bigg)\bigg\}\label{eq:exnum}
\end{align}
up to the linear order of $g$, where we introduced $C_{FP}:= -i[ F,\, P] = -i( F P-P F )$ and  $A_{FP}:=\{ P,\,  F\} =  P F +  F P$.  In particular, if we put $F = I_{\mathcal{P}}$ (the identity operator) above, we have 
\begin{align}
{\langle\Psi_f|  \Psi_f\rangle} \simeq \vert {\langle \psi_f| \psi_i\rangle} \vert^2 \bigg(
{\langle\Psi_i| \Psi_i\rangle} 
+ 2gt\,{\rm Im}\, A_w {\langle\Psi_i|}  P {|\Psi_i\rangle}\bigg).\label{eq:exden}
\end{align}
Denoting the expectation values of an observable $X$ evaluated for the final and the initial states of the probe by
\begin{align}
{\langle X\rangle}_f := \frac{{\langle \Psi_f|} X {|\Psi_f\rangle}}{{\langle\Psi_f|\Psi_f\rangle}}, \qquad {\langle  X \rangle}_i := \frac{{\langle\Psi_i|} X {|\Psi_i\rangle}}{{\langle\Psi_i|\Psi_i\rangle}},
\label{eq:defexp}
\end{align}
respectively, we find, from Eqs.~\eqref{eq:exnum} and \eqref{eq:exden},  
\begin{align}
{\langle  F \rangle}_f 
&\simeq {\langle F \rangle}_i + gt\bigg\{\,{\rm Re}\, A_w {\langle C_{FP}\rangle }_i + \,{\rm Im}\, A_w\bigg( {\langle A_{FP} \rangle}_i -2 {\langle  P \rangle}_i {\langle  F \rangle}_i\bigg)\bigg\}
\label{eq:expect}
\end{align}
up to the linear order of $g$.
We thus see that the shift of the pointer represented by $ F$ is
\begin{align}
\Delta F := {\langle  F \rangle}_f  - {\langle  F \rangle}_i \simeq  gt\bigg( {\rm Re}\, A_w {\langle C_{FP} \rangle}_i + 2\,{\rm Im}\, A_w {\rm Cov}(P,\, F)_i\bigg),
\label{eq:shift}
\end{align}
with 
\begin{align}
{\rm Cov}(P,\, F)_i := \frac{1}{2}{\langle\{P,\, F\} \rangle}_i -{\langle P \rangle}_i 
{\langle F \rangle}_i\label{eq:
covariance} 
\end{align}
being the (quantum) covariance for the observables $ P$ and $ F$ under the state ${|\Psi_i\rangle}$.
In particular, if we choose $ F = P$, the shift \eqref{eq:shift} reduces to~\cite{lee2013uncertainty} 
\begin{align}
\Delta P \simeq  2gt\,{\rm Im}\, A_w {\rm Var}( P)_i,\label{eq:shiftP}
\end{align}
where
\begin{align}
{\rm Var}( P)_i := {\langle P^2\rangle}_i - {\langle P \rangle}_i^2\label{eq:variance}
\end{align}
is the variance of $ P$ under the state ${|\Psi_i\rangle}$.  
If, instead, we choose $ F =  Q$ for which $[ Q,\,  P] =i$ (the observable canonically conjugate to $P$), then we have
\begin{align}
\Delta Q \simeq  gt\bigg( {\rm Re}\, A_w + 2\, {\rm Im}\, A_w {\rm Cov}(P,\,  Q)_i\bigg) ,
\label{eq:shiftQ}
\end{align}
  This implies that by choosing two distinct observables, one can obtain both the real and imaginary parts of the weak value $A_w$.  This provides the basics of the {\it indirect measurement} discussed in the literature.  


\subsection{Direct measurement}
\label{sec:Dir}
%
A special situation arises when the Hamiltonian of the system ${\cal S}$ is proportional solely to an observable of the system,
\begin{align}
 H = g A,
\label{eq:hamdirect}
\end{align}
where $g$ is a constant assumed to be small as before.  One is then allowed to dispense with the probe ${\cal P}$ to measure the weak value $A_w$~\cite{PhysRevLett.111.023604,PhysRevA.101.042117,RevModPhys.86.307,PhysRevLett.111.240402,Ueda:2023xoj}.  Indeed, one can obtain $A_w$ by directly looking at the transition probability 
\begin{align}
{\rm Pr}_{i \to f}(\epsilon) := \frac{\vert {\langle\psi_f|} e^{-i H t}  {|\psi_i\rangle} \vert^2}{\vert {\langle\psi_i| \psi_i\rangle}\vert} 
= \frac{\vert {\langle\psi_f|} e^{-i\epsilon  A}  {|\psi_i\rangle} \vert^2}{\vert {\langle \psi_i| \psi_i\rangle}\vert} ,
\label{eq:trprob}
\end{align}
where we put $\epsilon := gt$ for the overall constant, and the denominator (which is unnecessary when the initial state is normalized) is inserted to ensure the initial condition ${\rm Pr}_{i \to f}(0) = {\vert {\langle\psi_f| \psi_i\rangle}\vert^2}/{\vert {\langle \psi_i|\psi_i\rangle}\vert}$.
In this case, the transition probability is obtained by selecting the state ${|\psi_f\rangle}$ in the system ${\cal S}$ after the evolution time $t$, and thus ${|\psi_f\rangle}$ serves as the post-selected state.
Note that the transition probability \eqref{eq:trprob} may be regarded as a special case of measurement in the standard procedure without post-selection by means of the observable $ F =  E_f$ with 
\begin{align}
 E_f := {|\psi_f\rangle}{\langle\psi_f|},
\label{eq:projop}
\end{align}
which is a projection operator in ${\cal S}$ (in contrast to the observable in ${\cal P}$ discussed above) associated with the post-selection. Now, a procedure analogous to Eq.~\eqref{eq:psstwo} gives
\begin{align}
{\langle\psi_f|} e^{-i\epsilon  A}  {|\psi_i\rangle} \simeq {\langle \psi_f|} I_{\mathcal{S}} -i\epsilon  A  {|\psi_i\rangle} = {\langle\psi_f|\psi_i\rangle}\left(1-i\epsilon A_w \right).
\label{eq:pssthree}
\end{align}
Upon plugging this into Eq.~\eqref{eq:trprob}, one obtains
\begin{align}
{\rm Pr}_{i \to f}(\epsilon) = {\rm Pr}_{i \to f}(0)\left(1+i\epsilon A_w^* \right)\left(1-i\epsilon A_w \right) \simeq  {\rm Pr}_{i \to f}(0)\left(1 + 2\epsilon\,{\rm Im}\, A_w\right),\label{eq:wvdir}
\end{align}
up to the linear order of $\epsilon$.   We thus learn that the imaginary part ${\rm Im}\, A_w$ of the weak value can be found from 
the behavior of the transition amplitude~\cite{Denkmayr2014,hosten2008observation,PhysRevLett.102.173601,Song:2023uht,zhang2015precision,qiu2017precision}, or more explicitly, we have
\begin{align}
{\rm Im}\, A_w = \frac{1}{2\,{\rm Pr}_{i \to f}(0)}\frac{d}{d\epsilon}{\rm Pr}_{i \to f}(\epsilon)\biggr\vert_{\epsilon = 0} 
= \frac{1}{2}\frac{d}{d\epsilon}\ln {\rm Pr}_{i \to f}(\epsilon)\biggr\vert_{\epsilon = 0}.
\label{eq:imaw}
\end{align}

Since no probe $\cal P$ is used here, this measurement scheme is called   
{\it  direct measurement} of the weak value.  Although this is much simpler than the previous indirect measurement, the outcome \eqref{eq:imaw} shows a drawback of this scheme, that is, the real part ${\rm Re}\, A_w$ cannot be obtained here.  Clearly, this is due to the fact that the imaginary part ${\rm Im}\, A_w$ influences the modulus of the amplitude whereas the real part ${\rm Re}\, A_w$ affects its phase shift which is irrelevant to the probability.  
One possible remedy of this is to allow the parameter $\epsilon$ to be complex, which may arise when some non-unitary time evolution with a non self-adjoint Hamiltonian $H \ne  H^{\dagger}$ is considered.  Indeed, such a quantum system, called \lq non-Hermitian quantum mechanics\rq, has been studied intensively in recent years to accommodate a variety of physical situations (see, {\it e.g.}, Refs.~\cite{PhysRevLett.77.570, Ashida02072020}).  In that event, writing $\epsilon  = \epsilon_r + i \epsilon_i$ according to its real and imaginary parts, one obtains, instead of Eq.~\eqref{eq:wvdir}, 
\begin{align}
{\rm Pr}_{i \to f}(\epsilon) \simeq {\rm Pr}_{i \to f}(0)\left(1 + 2\epsilon_r\,{\rm Im}\, A_w + 2\epsilon_i\,{\rm Re}\, A_w\right),
\label{eq:wvdircomplex}
\end{align}
leading to
\begin{align}
{\rm Im}\, A_w = \frac{1}{2}\frac{d}{d\epsilon_r}\ln {\rm Pr}_{i \to f}(\epsilon)\biggr\vert_{\epsilon = 0}, \qquad 
{\rm Re}\, A_w = \frac{1}{2}\frac{d}{d\epsilon_i}\ln {\rm Pr}_{i \to f}(\epsilon)\biggr\vert_{\epsilon = 0},
\label{eq:imawcomplex}
\end{align}
respectively.  A simplest example of such cases is provided by, for instance, the system that exhibits a decay of particles (states), and another will arise when one wishes to focus on a subset of states for possible transitions admitted in the system.

The forgoing argument suggests that in the direct measurement scheme the weak value $A_w$, which is complex in general, is more naturally associated with the transition amplitude than the transition probability.  In fact, from Eq.~\eqref{eq:pssthree} the amplitude reads
\begin{align}
{\rm Amp}_{i \to f}(\epsilon) := \frac{{\langle\psi_f|} e^{-igt  H}  {|\psi_i\rangle} } {\sqrt{\vert{\langle\psi_i| \psi_i\rangle}\vert}}
\simeq {\rm Amp}_{i \to f}(0) \left(1-i\epsilon A_w \right),
\label{eq:tramp}
\end{align}
which indicates that the weak value admits the compact formula,
\begin{align}
A_w = i\frac{d}{d\epsilon}\ln {\rm Amp}_{i \to f}(\epsilon)\biggr\vert_{\epsilon = 0},
\label{eq:wvexp}
\end{align}
without rendering the parameter $\epsilon$ complex.  In particular, taking the imaginary part gives
\begin{align}
{\rm Im}\, A_w = {\rm Re}\, \frac{d}{d\epsilon}\ln {\rm Amp}_{i \to f}(\epsilon)\biggr\vert_{\epsilon = 0}
= \frac{1}{2}\frac{d}{d\epsilon}\ln {\rm Pr}_{i \to f}(\epsilon)\biggr\vert_{\epsilon = 0},
\label{eq:wvexpim}
\end{align}
assuring the result \eqref{eq:imaw}.
From this we learn that the weak value represents the susceptibility of the time evolution at $\epsilon = 0$ against the operator $A$ which constitutes the Hamiltonian.  Even though the transition amplitude \eqref{eq:tramp} is not directly observable, the above relation \eqref{eq:wvexp} suggests that the weak value is intrinsically quantum and fundamental in describing the transition of quantum states.


\subsection{Indirect and direct measurements embedded in a unified framework}
\label{sec:unif}
%
It is possible to embed both the indirect and direct measurements formally into a unified framework consisting of both $\cal S$ and $\cal P$.  To see this, we revisit the indirect measurement mentioned earlier and consider the observable, 
\begin{align}
{\cal F} := {|\psi_f\rangle}{\langle\psi_f|}\otimes  F =   E_f \otimes  F,
\label{eq:obscomb}
\end{align}
of the combined system, where $ E_f$ is the projection operator onto the post-selected state \eqref{eq:projop} and $F$ is an arbitrary observable of the probe $\cal P$.  The initial and the final states of the combined system for which our measurement is to be implemented are
\begin{align}
{|\Lambda_f\rangle} := e^{-i H t}  {|\Lambda_i\rangle}, \qquad {|\Lambda_i\rangle} := {|\psi_i\otimes\Psi_i\rangle},
\label{eq:stcomb}
\end{align}
as given in Eq.~\eqref{eq:tev} with the Hamiltonian \eqref{eq:ham}.  In this scheme, the expectation value of the observable ${\cal F}$ for the final combined state, for instance, is given by
\begin{align}
{\langle{\cal F} \rangle}_f := \frac{{\langle\Lambda_f|} {\cal F} {|\Lambda_f\rangle}}{{\langle\Lambda_f|\Lambda_f\rangle}}.
\label{eq:expcomb}
\end{align}
Recalling Eq.~\eqref{eq:pss}, {\it i.e.},
\begin{align}
 {|\Psi_f\rangle} = {\langle\psi_f|}{\Lambda_f\rangle},
\label{eq:crrel}
\end{align}
we have
\begin{align}
{\langle\Lambda_f|} {\cal F} {|\Lambda_f\rangle} = {\langle\psi_i\otimes\Psi_i|} e^{i H t} \left( {|\psi_f\rangle}{\langle\psi_f|}\otimes  F\right) e^{-i H t}  {|\psi_i\otimes\Psi_i\rangle}
= {\langle\Psi_f|}  F {|\Psi_f\rangle}.
\label{eq:identcomb}
\end{align}
Similarly, we find
\begin{align}
 {\langle \Lambda_i|} {\cal F} {|\Lambda_i\rangle} = {\langle\psi_i\otimes\Psi_i|}  \left( {|\psi_f\rangle}{\langle\psi_f|}\otimes  F\right) {|\psi_i\otimes\Psi_i\rangle}
= \vert{\langle\psi_f|\psi_i\rangle}\vert^2 {\langle\Psi_i|}  F {|\Psi_i\rangle},
\label{eq:identcombini}
\end{align}
where we notice that the operator ${\cal F}$ effectively implements the post-selection despite being measured for the initial state.
We also note from \eqref{eq:identcomb} that the numerator of the expectation value \eqref{eq:expcomb} yields precisely the value of the numerator of the expectation value obtained when $ F$ is measured for the state ${|\Psi_f\rangle}$ in the probe $\cal P$.   However, the denominator of the expectation value in Eq.~\eqref{eq:defexp} does not, since Eq.~\eqref{eq:crrel} implies that
\begin{align}
{\langle\Lambda_f|\Lambda_f\rangle} \ne {\langle\Lambda_f|}  E_f\otimes I_{\mathcal{P}} {|\Lambda_f\rangle} = {\langle\Psi_f|\Psi_f\rangle},
\label{eq:difden}
\end{align}
even though it is independent of the time evolution
${\langle\Lambda_f|\Lambda_f\rangle} = {\langle\Lambda_i|\Lambda_i\rangle}$. 
It follows that the expectation value of the combined system  \eqref{eq:expcomb} turns out to be something different in relation to the weak value from those obtained in the indirect measurement discussed before.  

More explicitly, we observe that the shift in the expectation value due to the time evolution and the post-selection can be evaluated from Eqs.~\eqref{eq:expcomb} and \eqref{eq:identcomb} as
\begin{align} 
\Delta {\cal F} := {\langle{\cal F} \rangle}_f  - {\langle {\cal F} \rangle}_i 
&= \frac{{\langle\Lambda_f|}  F {|\Lambda_f\rangle}}{{\langle\Lambda_f|\Lambda_f\rangle}} -
\frac{\vert{\langle\psi_f|\psi_i\rangle}\vert^2 {\langle\Psi_i|}  F {|\Psi_i\rangle}}{{\langle\Lambda_i|\Lambda_i\rangle}}\notag
\\
&= \frac{{\langle\Psi_f|}  F {|\Psi_f\rangle}}{{\langle\Lambda_f|\Lambda_f\rangle}} - {\langle E_f\rangle}_i {\langle  F\rangle}_i.
\label{eq:shiftcomb}
\end{align}
Comparing with Eq.~\eqref{eq:exnum}, we observe that here both the real and the imaginary parts of the weak value turn out to be rather convoluted.  Fortunately, the gap between the two may be filled by considering 
the ratio of expectation values,
\begin{align}
\frac{{\langle {\cal F} \rangle}_f}{\langle  E_f\otimes I_{\mathcal{P}}\rangle_f} = \frac{{\langle\Lambda_f|} {\cal F} {|\Lambda_f\rangle}}{{\langle\Lambda_f|\Lambda_f\rangle}}
\frac{{\langle\Lambda_f|\Lambda_f\rangle}}{{\langle\Lambda_f|}  E_f\otimes I_{\mathcal{P}} {|\Lambda_f\rangle}} = \frac{{\langle\Psi_f|}  F {|\Psi_f\rangle}}{{\langle\Psi_f|\Psi_f\rangle}} 
= {\langle  F \rangle}_f,
\label{eq:expratio}
\end{align}
where we have used Eqs.~\eqref{eq:identcomb} and \eqref{eq:difden}.  Indeed, since this coincides with the expectation value of the observable $ F$ in the indirect measurement,  combining with the similar ratio for the initial state, we find
\begin{align}
\frac{{\langle {\cal F} \rangle_f}}{{\langle  E_f\otimes I_{\mathcal{P}}\rangle}_f}  - \frac{{\langle {\cal F} \rangle}_i}{{\langle  E_f\otimes I_{\mathcal{P}}\rangle}_i} =  {\langle  F \rangle}_f -  {\langle  F \rangle}_i = \Delta F,
\label{eq:equiind}
\end{align}
which agrees with the shift \eqref{eq:shift}.  We therefore find that the indirect measurement of the weak value can be embedded in the present combined scheme once the extra procedure of taking the ratio \eqref{eq:expratio} is augmented.

Turning to the direct measurement, we recall the Hamiltonian \eqref{eq:hamdirect} which, in the combined system, takes the form 
$ H = g A \otimes I_{\mathcal{P}}$ and hence yields the unitary evolution, 
\begin{align}
e^{-i\epsilon  A \otimes I_{\mathcal{P}}} = e^{-i\epsilon  A} \otimes I_{\mathcal{P}}.
\label{eq:tevco}
\end{align}
Now, since from Eq.~\eqref{eq:crrel} we have 
\begin{align}
{\langle\Lambda_f|}  E_f\otimes I_{\mathcal{P}} {|\Lambda_f\rangle} 
&= {\langle\psi_i\otimes\Psi_i|}\, \left(e^{+i\epsilon  A} \otimes I_{\mathcal{P}}\right)  E_f\otimes I_{\mathcal{P}} \left(e^{-i\epsilon  A} \otimes I_{\mathcal{P}}\right) \,{|\psi_i\otimes\Psi_i\rangle} \notag
\\
&={\rm Pr}_{i \to f}(\epsilon) \vert {\langle\psi_i| \psi_i\rangle}\vert {\langle\Psi_i| \Psi_i\rangle},
\label{eq:equiind}
\end{align}
and also
\begin{align}
{\langle\Lambda_f|\Lambda_f\rangle} = {\langle\psi_i| \psi_i\rangle} {\langle\Psi_i|\Psi_i\rangle},
\label{eq:equiindden}
\end{align}
we arrive at
\begin{align}
{\langle  E_f\otimes I_{\mathcal{P}}\rangle}_f = \frac{{\langle\Lambda_f|}  E_f\otimes I_{\mathcal{P}} {|\Lambda_f\rangle}}{{\langle\Lambda_f|\Lambda_f\rangle}} 
= {\rm Pr}_{i \to f}(\epsilon).
\label{eq:equidir}
\end{align}
We thus find that the result of the direct measurement can also be obtained in the present scheme as a single expectation value.

To sum up, we have seen that there exists a unified framework in which both the indirect and the direct measurements can be embedded.  This scheme allows us to obtain exactly the same results for both the indirect and the direct measurements, provided that the ratio of the expectation value \eqref{eq:expratio} is used in the former case whereas the trivial observable $ F = I_{\mathcal{P}}$ is chosen in the latter case.

\section{Weak measurement in qubit systems}
\label{sec:weak_qubit}
For our purpose of reconsidering quantum resonance in the context of weak measurement, 
in this section we focus on the system $\mathcal{S}$ consisting of a qubit, {\it i.e.}, the system spanned by two independent states such as the spin states of the electron.  We analyze the process of weak measurement in two distinct schemes, direct and indirect, with some technical detail so that the results can be used later in our discussion of quantum resonance.
We first review the direct measurement along the line mentioned in  Ref.\cite{Ueda:2023xoj}
and extend it to the scheme of indirect measurement following the notation used there.

\subsection{Direct measurement}
\label{sec:direct}
In order to discuss a general measurement process in the system $\mathcal{S}$ of a qubit, one may consider the Hamiltonian,
\begin{align}
    H_{\rm dir}=H_0+V,
\label{eq:Ham1}
\end{align}
where $H_0$ is the \lq free\rq\ part and $V$ is the \lq interaction\rq\ part describing the measurement. 
In relation to Section~\ref{sec:Dir}, in Eq.~\eqref{eq:hamdirect}, $H_0=0$ and $V=gA$ are assumed.
Being a Hermitian $2 \times 2$ matrix, each of them can be parametrized as
\begin{align}
    H_0 &= h \left(n_0^{(h)}+ \vec{n}^{(h)}\cdot \vec{\sigma}\right),\qquad V = v\left(n_0^{(v)}+ \vec{n}^{(v)}\cdot \vec{\sigma}\right),
    \label{eq:h0V}
\end{align}
with real positive values $h, v$ and $\vec{\sigma}$ consisting of the Pauli matrices $\vec{\sigma} = (\sigma_1,\, \sigma_2,\, \sigma_3)$.
The terms with the coefficients $n_0^{(h)}$ and $n_0^{(v)}$ denote the components proportional to the $2 \times 2$ identity matrix corresponding to the identity operator $ I_{\mathcal{S}}$ (suppressed here for brevity), and
$\vec{n}^{(h)}\cdot \vec{\sigma}=\sum_{i=1}^3 n_{i}^{(h)} \sigma_i$ and $\vec{n}^{(v)}\cdot \vec{\sigma}=\sum_{i=1}^3 n_i^{(v)} \sigma_i$.
The vectors $\vec{n}^{(h)}=(n_1^{(h)},\, n_2^{(h)},\, n_3^{(h)})$ and $\vec{n}^{(v)}=(n_1^{(v)},\, n_2^{(v)},\, n_3^{(v)})$
are normalized such that $\vec{n}^{(h)}\cdot \vec{n}^{(h)\ast}=1$ and $\vec{n}^{(v)}\cdot \vec{n}^{(v)\ast}=1$.
The components $n_0^{(h)}, n_0^{(v)}$ and $n_i^{(h)}, n_i^{(v)}$ for $i=1,2,3$ may take complex values if the time evolution is governed by a non-Hermitian Hamiltonian. 

Let ${|\psi(t)\rangle}$ be the quantum state of the system at time $t$.
According to Eq.~\eqref{eq:Ham1}, the time evolution of the state from $t_0$ to $t_0 + t$ is given by
\begin{align}
{|\psi(t_0+t)\rangle}&=e^{-i H_{\rm dir}t}{|\psi_i\rangle},\label{eq:time1}
\end{align}
where ${|\psi_i\rangle}:={|\psi(t_0)\rangle}$.
The time evolution~\eqref{eq:time1} can be approximated, up to the first order of $v$, as
\begin{align}
    {|\psi(t_0+t)\rangle}&\simeq  e^{-ivn_0^{(v)}t}\left[
    e^{-i H_0 t}
    \left(1-ivt \frac{\vec{n}^{(h)}\cdot \vec{n}^{(v)}}{(\vec{n}^{(h)})^2}\sigma_{h} \right)
    -i \frac{v}{2h}[\sigma_{a},\, e^{-iH_0 t}]
    \right]{|\psi_i\rangle},
    \label{eq:time2}
\end{align}
by using the shorthands,
\begin{equation}
\sigma_{h}:=\vec{n}^{(h)}\cdot \vec{\sigma}, \qquad \sigma_{v}:=\vec{n}^{(v)}\cdot \vec{\sigma}, \qquad \sigma_{a}:=\vec{n}^{(a)}\cdot \vec{\sigma},
\end{equation}
with $\vec{n}^{(a)}=(n_1^{(a)},\, n_2^{(a)},\, n_3^{(a)})$ satisfying the normalization condition $\vec{n}^{(a)}\cdot \vec{n}^{(a)\ast}=1$, where $\sigma_{a}$ is chosen to fulfill the relation,
\begin{align}
[\sigma_{a},\, \sigma_{h}]=2i
\left(\sigma_{v}-\frac{\vec{n}^{(h)}\cdot \vec{n}^{(v)}}{(\vec{n}^{(h)})^2}\sigma_{h}\right),
\label{eq:condsa}
\end{align}
with $\sigma_{h}$ and $\sigma_{v}$.

After the time evolution described by Eq.~\eqref{eq:time2}, we perform the post-selection by a state ${|\psi_f\rangle}$ arbitrarily chosen in the system $\mathcal{S}$.
Under this, the transition amplitude from ${|\psi_i\rangle}$ to ${|\psi_f\rangle}$ is found to be
\begin{align}
    {\langle \psi_f|\psi(t_0+t)\rangle}&= {\langle \psi_f|}e^{-i H_{\rm dir} t}{|\psi_i\rangle}\notag
    \\
&\simeq e^{-ivn_0^{(v)}t}{\langle \psi_f |e^{-i H_0 t}|\psi_i\rangle}
    \left(
    1-i vt \frac{\vec{n}^{(h)}\cdot \vec{n}^{(v)}}{(\vec{n}^{(h)})^2}\sigma^W_{h}
-i\frac{v}{2h}\left(\sigma^W_{a,L}-\sigma^W_{a,R}\right)
    \right)
    ,\label{eq:amp1}
\end{align}
where 
\begin{align}
    \sigma_{h}^W:&=\frac{{\langle \psi_f|}\sigma_{h} e^{-iH_0 t}{|\psi_i\rangle}}{{\langle \psi_f|}e^{-iH_0 t}{|\psi_i\rangle}}= \frac{{\langle \psi_f|}e^{-iH_0t}\sigma_{h} {|\psi_i\rangle}}{{\langle \psi_f|}e^{-iH_0t}{|\psi_i\rangle}},
    \\
    \sigma^W_{{a},L}:&=\frac{{\langle \psi_f|\sigma_{a}e^{-iH_0t}|\psi_i\rangle}}{{\langle \psi_f|e^{-iH_0t}|\psi_i\rangle}},\qquad \sigma^W_{{a},R}:=\frac{{\langle \psi_f|e^{-iH_0t}\sigma_{a}|\psi_i\rangle}}{{\langle \psi_f|e^{-iH_0t}|\psi_i\rangle}},\label{eq:weakI}
\end{align}
represent the weak values associated with $\sigma_h$ and $\sigma_a$, respectively.  The lower subscripts $L$ and $R$ are introduced to indicate when the weak value is measured, either at the initial or the final time.  
From these, the transition probability from ${|\psi_i\rangle}$ to ${|\psi_f\rangle}$ becomes
\begin{align}
    {\rm Pr}_{i\to f}\left(v\right)&:=\left|{\langle \psi_f|\psi(t_0+t)\rangle}\right|^2\notag
    \\
    &\simeq {\rm Pr}_{i\to f}\left(0\right)e^{2vt\,{\rm Im}\,n_0^{(v)}}
    \Bigg[
    1+2vt \bigg\{
    {\rm Im}\,\left(\frac{\vec{n}^{(h)}\cdot \vec{n}^{(v)}}{(\vec{n}^{(h)})^2}\right){\rm Re}\,\sigma^W_{h}\notag
    \\
    &\qquad\qquad+{\rm Re}\,\left(\frac{\vec{n}^{(h)}\cdot \vec{n}^{(v)}}{(\vec{n}^{(h)})^2}\right){\rm Im}\,\sigma^W_{h}
    \bigg\}+\frac{v}{h} \left({\rm Im}\,\sigma^W_{{a},L}-{\rm Im}\,\sigma^W_{{a},R}\right)
    \Bigg],\label{eq:pr1}
\end{align}
with ${\rm Pr}_{i\to f}(0)=\left|{\langle \psi_f |e^{-iH_0 t}|\psi_i\rangle}\right|^2$.
Below we examine the weak measurement in this setup in more detail, focusing on two specific cases: $\vec{n}^{(h)}\times \vec{n}^{(v)}=0$ and $\vec{n}^{(h)}\cdot \vec{n}^{(v)}=0$.

\subsubsection{Commutative case
}
In the special case where $\vec{n}^{(h)} \times \vec{n}^{(v)} = 0$, {\it i.e.}, $\vec{n}^{(h)} \propto \vec{n}^{(v)}$, the two operators $H_0$ and $V$ commute, $[H_0,\, V] = 0$.
As will be shown later, the Ramsey resonance corresponds to this case.
Assuming $n_0^{(v)} = 0$, the time evolution described by Eq.~\eqref{eq:time2} is, up to the first order in $v$, given by
\begin{align}
    {|\psi(t_0+t)\rangle}
    &\simeq e^{-i H_0 t} \left(1-iv t \sigma_v\right){|\psi_i\rangle},
    \label{eq:timeI1}
\end{align}
%
%
and hence
\begin{align}
    {\langle \psi_f|\psi(t_0+t)\rangle}\simeq {\langle \psi_f|}e^{-iH_0t}{|\psi_i\rangle} \left(1-iv t \sigma_{v}^W\right)\label{eq:ampI}.
\end{align}
%
%
%
Upon substituting Eq.~\eqref{eq:ampI} into Eq.~\eqref{eq:pr1}, we obtain
\begin{align}
    {\rm Pr}_{i\to f}\left(v\right)
    &\simeq {\rm Pr}_{i\to f}\left(0\right) \left(1+2vt\,{\rm Im}\, \sigma_{v}^W\right).\label{eq:PrI}
\end{align}
This corresponds to Eq.~\eqref{eq:wvdir} and represents the effect of the direct measurement with post-selection up to the first order in $v$, {\it i.e.,} the direct weak measurement contribution to the transition probability.
From Eq.~\eqref{eq:PrI}, the imaginary part of the weak value of $\sigma_v$ can be written as
\begin{align}
    {\rm Im}\,\sigma_{v}^W =
    \frac{1}{2{\rm Pr}_{i\to f}\left(0\right)}\frac{d{\rm Pr}_{i\to f}\left(v\right)}{d(vt)}\bigg|_{v=0}.\label{eq:weakI2}
\end{align}
This formula shows that the weak value ${\rm Im}\,\sigma_{v}^W$ represents the susceptibility with respect to the interaction $V$, where 
the parameter $vt$ appearing in Eq.~\eqref{eq:weakI2} is identified as the measurement strength. All of these are consistent with the result Eq.~\eqref{eq:imaw} mentioned before.  
Our calculations, performed to the first order in $vt$, are valid for weak measurements under the condition $vt \ll 1$.

\subsubsection{Non-commutative case
}
Another special case arises when $\vec{n}^{(h)} \cdot \vec{n}^{(v)} = 0$ holds, for which the two operators $H_0$ and $V$ do not commute, $[H_0,\, V] \neq 0$.
As will be shown later, the Rabi resonance falls into this category.
Assuming $n_0^{(v)} = 0$, the time evolution described by Eq.~\eqref{eq:time2} is, up to the first order in $v$, given by
\begin{align}
    {|\psi(t_0+t)\rangle}\simeq \left(e^{-i H_0t}-i\frac{v}{2h} [\sigma_{a},\,e^{-iH_0t}]\right){|\psi_i\rangle},
    \label{eq:timeI}
\end{align}
which yields
\begin{align}
    {\langle \psi_f|\psi(t_0+t)\rangle}\simeq {\langle \psi_f|e^{-iH_0t}|\psi_i\rangle}
    \left(1-i\frac{v}{2h}\left( \sigma_{{a},L}^W- \sigma_{a,R}^W \right)\right).
    \label{eq:ampII}
\end{align}
Upon substituting Eq.~\eqref{eq:ampII} into Eq.~\eqref{eq:pr1}, we obtain
\begin{align}
    {\rm Pr}_{i\to f}\left(v\right)
    &\simeq {\rm Pr}_{i\to f}\left(0\right) 
    \left(1+\frac{v}{h} \left({\rm Im}\,\sigma^W_{{a},L}-{\rm Im}\,\sigma^W_{{a},R}\right)\right),\label{eq:PrII}
\end{align}
which implies
\begin{align}
    {\rm Im}\,\sigma^W_{a,L}-{\rm Im}\,\sigma^W_{a,R}=
    \frac{1}{{\rm Pr}_{i\to f}\left(0\right)}\frac{d{\rm Pr}_{i\to f}\left(v\right)}{d(v/h)}\bigg|_{v=0}.\label{eq:weakII}
\end{align}
The ratio of parameters $v/h$ appearing in Eq.~\eqref{eq:weakII} may be identified as the measurement strength of the direct weak measurement.
This outcome shows that the present weak measurement enables us to make a distinction between ${\rm Im}\,\sigma^W_{a,L}$ and ${\rm Im}\,\sigma^W_{a,R}$.  Our arguments hold as long as the condition $v/h \ll 1$ is fulfilled.
%
%
%

%
%

\subsection{Indirect measurement}
Next, we turn our attention to the indirect measurement scheme corresponding to the setup discussed in the direct measurement scheme above.
To this end, we consider a quantum system composed of the qubit system $\mathcal{S}$ and a probe system $\mathcal{P}$ described by the following Hamiltonian,
\begin{align}
    H_{\rm ind}=H_0\otimes I_{\mathcal{P}} +V\otimes P,\label{eq:Ham2}
\end{align}
where $H_0$ and $V$ are defined in Eq.~\eqref{eq:h0V}, and $P$ is an operator acting on the probe system $\mathcal{P}$.
Again, in relation to Section~\ref{sec:Ind}, in Eq.~\eqref{eq:ham}, $H_0=0$ and $V=g A$ are assumed.
Setting $P=I_P$, Eq.~\eqref{eq:Ham2} reduces to Eq.~\eqref{eq:Ham1} in the context of the unified framework provided in Section~\ref{sec:unif}.
According to Eq.~\eqref{eq:Ham2}, the time evolution of the quantum state of the entire system $\mathcal{S}\otimes\mathcal{P}$ between $t_0$ and $t_0 + t$ is given by
\begin{align}
	{|\Lambda_f\rangle}&=e^{-i H_{\rm ind} t} {|\Lambda_i\rangle}\notag
    \\
 &\simeq e^{-i v n_0^{(v)} t \left(I_{\mathcal{S}}\otimes P\right) }
	\Bigg[
	e^{-i \left(H_0\otimes I_{\mathcal{P}}\right) t} \left(I_{\mathcal{S}}\otimes I_{\mathcal{P}}-i vt \frac{\vec{n}^{(h)}\cdot \vec{n}^{(v)}}{(\vec{n}^{(h)})^2}\sigma_h\otimes P\right)\notag
 \\
	&\quad\quad\quad\quad\quad\quad\quad\quad\qquad\qquad\qquad\qquad-i \frac{v}{2h}[\sigma_a,\, e^{-i H_0 t}]\otimes P
	\Bigg] {|\psi_i\otimes \Psi_i\rangle},\label{eq:time3}
\end{align}
where we take the initial state of the combined system to be ${|\Lambda_i\rangle} := {|\psi_i \otimes \Psi_i\rangle}$ which is the direct product of the system state ${|\psi_i\rangle}$ and the probe state ${|\Psi_i\rangle}$.
As in the direct weak measurement, after the time evolution given in Eq.~\eqref{eq:time3}, we perform the post-selection on the qubit system $\mathcal{S}$, focusing on the final state ${|\psi_f\rangle}$ at the end of the measurement.
The quantum state of the probe system $\mathcal{P}$ after the post-selection is approximated, up to the first order in $v$, as
\begin{align}
	{|\Psi_f\rangle}&={\langle \psi_f | \Lambda_f\rangle}\notag
	\\
	&\simeq {\langle \psi_f|e^{-i H_0 t} |\psi_i\rangle}\bigg[
	I_{\mathcal{P}} -i v n_0^{(v)}t P-iv t \frac{\vec{n}^{(h)}\cdot \vec{n}^{(v)}}{(\vec{n}^{(h)})^2}\sigma^W_h P
-i \frac{v}{2h} \left(\sigma^W_{a,L}-\sigma^W_{a,R}\right) P
	\bigg]{|\Psi_i\rangle}.\label{eq:Psif_qubit}
\end{align}

From these expressions, we find the expectation value of the observable $F$ in the probe system $\mathcal{P}$ after the post-selection,
\begin{align}
    {\langle \Psi_f|F|\Psi_f\rangle}&\simeq \left|{\langle \psi_f|e^{-i H_0 t}|\psi_i\rangle} \right|^2\notag
    \\
    &\times
    \Bigg[
	{\langle \Psi_i|F|\Psi_i\rangle}+\bigg\{vt \left({\rm Re}\, n_0^{(v)}
 +
 {\rm Re}\, \left(\frac{\vec{n}^{(h)}\cdot\vec{n}^{(v)}}{(\vec{n}^{(h)})^2}\right) {\rm Re}\,\sigma^W_h
 -{\rm Im}\,\left(\frac{\vec{n}^{(h)}\cdot\vec{n}^{(v)}}{(\vec{n}^{(h)})^2}\right){\rm Im}\,\sigma^W_h
 \right)\notag
\\
&+\frac{v}{2h}{\rm Re}\,\left( \sigma_{a,L}^W-\sigma_{a,R}^W\right)
 \bigg\}{\langle \Psi_i|C_{FP}|\Psi_i\rangle}\notag
 \\
 &+\bigg\{vt \left({\rm Im}\, n_0^{(v)}
 +
 {\rm Re}\, \left(\frac{\vec{n}^{(h)}\cdot\vec{n}^{(v)}}{(\vec{n}^{(h)})^2}\right) {\rm Im}\,\sigma^W_h
 +{\rm Im}\,\left(\frac{\vec{n}^{(h)}\cdot\vec{n}^{(v)}}{(\vec{n}^{(h)})^2}\right){\rm Re}\,\sigma^W_h
 \right)\notag
\\
&+\frac{v}{2h}{\rm Im}\,\left( \sigma_{a,L}^W-\sigma_{a,R}^W\right)
 \bigg\}{\langle \Psi_i|A_{FP}|\Psi_i\rangle}\Bigg].\label{eq:X_qubit}
\end{align}
%
%
Eq.~\eqref{eq:X_qubit} corresponds to Eq.~\eqref{eq:exnum} for the qubit system, and the weak values represent the shift in the expectation value of $F$ induced by the measurement process involving the interaction term in Eq.~\eqref{eq:Ham2}.
Analogously to the derivation of Eq.~\eqref{eq:shift}, the shift of the pointer $F$ is obtained from Eq.~\eqref{eq:X_qubit} as
\begin{align}
    \Delta F&= {\langle F\rangle}_f- {\langle F\rangle}_i\notag
    \\
    &\simeq  \bigg\{vt \left({\rm Re}\, n_0^{(v)}
 +
 {\rm Re}\, \left(\frac{\vec{n}^{(h)}\cdot\vec{n}^{(v)}}{(\vec{n}^{(h)})^2}\right) {\rm Re}\,\sigma^W_h
 -{\rm Im}\,\left(\frac{\vec{n}^{(h)}\cdot\vec{n}^{(v)}}{(\vec{n}^{(h)})^2}\right){\rm Im}\,\sigma^W_h
 \right)\notag
\\
&+\frac{v}{2h}{\rm Re}\,\left( \sigma_{a,L}^W-\sigma_{a,R}^W\right)
 \bigg\}{\langle C_{FP}\rangle}_i\notag
 \\
 &+2\bigg\{vt \left({\rm Im}\, n_0^{(v)}
 +
 {\rm Re}\, \left(\frac{\vec{n}^{(h)}\cdot\vec{n}^{(v)}}{(\vec{n}^{(h)})^2}\right) {\rm Im}\,\sigma^W_h
 +{\rm Im}\,\left(\frac{\vec{n}^{(h)}\cdot\vec{n}^{(v)}}{(\vec{n}^{(h)})^2}\right){\rm Re}\,\sigma^W_h
 \right)\notag
\\
&+\frac{v}{2h}{\rm Im}\,\left( \sigma_{a,L}^W-\sigma_{a,R}^W\right)
 \bigg\} {\rm Cov}\left(F,\,P\right)_i.\label{eq:sfit_F}
\end{align}
For $F=P$, the shift~\eqref{eq:sfit_F} reduces to
\begin{align}
    \Delta P&\simeq 2\bigg[vt \left({\rm Im}\, n_0^{(v)}
 +
 {\rm Re}\, \left(\frac{\vec{n}^{(h)}\cdot\vec{n}^{(v)}}{(\vec{n}^{(h)})^2}\right) {\rm Im}\,\sigma^W_h
 +{\rm Im}\,\left(\frac{\vec{n}^{(h)}\cdot\vec{n}^{(v)}}{(\vec{n}^{(h)})^2}\right){\rm Re}\,\sigma^W_h
 \right)\notag
\\
&\qquad\qquad +\frac{v}{2h}{\rm Im}\,\left( \sigma_{a,L}^W-\sigma_{a,R}^W\right)
 \bigg] {\rm Var}(P)_i.\label{eq:DelP_ind}
\end{align}
Instead, for $F=Q$ with $[Q,\, P]=i$, we obtain
\begin{align}
    \Delta Q&={\langle Q\rangle}_f-{\langle Q\rangle}_i\notag
    \\
    &\simeq \bigg\{vt \left({\rm Re}\, n_0^{(v)}
 +
 {\rm Re}\, \left(\frac{\vec{n}^{(h)}\cdot\vec{n}^{(v)}}{(\vec{n}^{(h)})^2}\right) {\rm Re}\,\sigma^W_h
 -{\rm Im}\,\left(\frac{\vec{n}^{(h)}\cdot\vec{n}^{(v)}}{(\vec{n}^{(h)})^2}\right){\rm Im}\,\sigma^W_h
 \right)\notag
\\
&\qquad\qquad\qquad+\frac{v}{2h}{\rm Re}\,\left( \sigma_{a,L}^W-\sigma_{a,R}^W\right)
 \bigg\}\notag
 \\
 &+2\bigg\{vt \left({\rm Im}\, n_0^{(v)}
 +
 {\rm Re}\, \left(\frac{\vec{n}^{(h)}\cdot\vec{n}^{(v)}}{(\vec{n}^{(h)})^2}\right) {\rm Im}\,\sigma^W_h
 +{\rm Im}\,\left(\frac{\vec{n}^{(h)}\cdot\vec{n}^{(v)}}{(\vec{n}^{(h)})^2}\right){\rm Re}\,\sigma^W_h
 \right)\notag
\\
&\qquad\qquad\qquad+\frac{v}{2h}{\rm Im}\,\left( \sigma_{a,L}^W-\sigma_{a,R}^W\right)
 \Bigg\} {\rm Cov}\left(P,\,Q\right)_i.\label{eq:DelQ_ind}
\end{align}
As will be discussed in more detail for two representative cases, $\vec{n}^{(h)}\times \vec{n}^{(v)}=0$ and $\vec{n}^{(h)}\cdot \vec{n}^{(v)}=0$ both under the condition $n_0^{(v)}=0$, the weak values can be extracted by combining Eqs.~\eqref{eq:DelP_ind} and \eqref{eq:DelQ_ind}.

\subsubsection{Commutative case
}
For the commutative case $\vec{n}^{(h)}\times \vec{n}^{(v)}=0$ for which $[H_0,\,V]=0$ holds, the time evolution of Eq.~\eqref{eq:time3} reads 
\begin{align}
	{|\Lambda_f\rangle}=e^{-i H_{\rm ind} t} {|\psi_i\otimes \Psi_i\rangle}\simeq e^{-i \left(H_0\otimes I_{\mathcal{P}}\right) t} \left(I_{\mathcal{S}}\otimes I_{\mathcal{P}}-i vt \sigma_v \otimes P\right){|\psi_i\otimes \Psi_i\rangle}.
\end{align}
The post-selection then renders the state of the probe system $\mathcal{P}$ to be
\begin{align}
{|\Psi_f\rangle}&= {\langle \psi_f|}e^{-i H_{\rm ind} t} {|\psi_i\otimes \Psi_i\rangle}\simeq {\langle \psi_f|e^{-i H_0 t}|\psi_i\rangle}
\left(
I_{\mathcal{P}}-ivt \sigma^W_h P
\right) {|\Psi_i\rangle},\label{eq:Psif_com}
\end{align}
where $\sigma_h=\sigma_v$.
From Eq.~\eqref{eq:Psif_com}, the expectation value of $F$ after the post-selection is evaluated, up to the first order of $v$, as
\begin{align}
    {\langle \Psi_f|F|\Psi_f\rangle}
    &\simeq \left|{\langle \psi_f|e^{-iH_0 t}|\psi_i\rangle}\right|^2
    \notag
    \\
    &\times \bigg[
    {\langle \Psi_i|F|\Psi_i\rangle}+vt \left(
    {\rm Re}\, \sigma^W_h {\langle \Psi_i|C_{FP}|\Psi_i\rangle}
    +
    {\rm Im}\, \sigma^W_h 
    {\langle \Psi_i|A_{FP}|\Psi_i\rangle}
    \right)
    \bigg].
\end{align}
%
From these, we obtain
\begin{align}
    \Delta F &=
    {\langle F\rangle}_f-{\langle F\rangle}_i\simeq
    vt \bigg[
    {\rm Re}\, \sigma^W_h {\langle C_{FP}\rangle}_i
    +2{\rm Im}\, \sigma^W_h {\rm Cov}\left(F,\,P\right)_i
    \bigg].\label{eq:DelF_com}
\end{align}
For $F=P$ and $Q$, Eq.~\eqref{eq:DelF_com} yields,
\begin{align}
    \Delta P&\simeq 2vt\, {\rm Im}\, \sigma^W_h {\rm Var}\left(P\right)_i,\label{eq:Pexp_com}
    \\
    \Delta Q&\simeq vt \left(
    {\rm Re}\, \sigma^W_h
    +2{\rm Im}\, \sigma^W_h {\rm Cov}\left(P,\,Q\right)_i
    \right).\label{eq:Xexp_com}
\end{align}
Upon combining Eqs.~\eqref{eq:Xexp_com} and \eqref{eq:Pexp_com}, real and imaginary parts of the weak value $\sigma^W_h$ are extracted as 
\begin{align}
{\rm Im}\, \sigma^W_h &\simeq \frac{1}{2 {\rm Var}\left(P\right)_i} \frac{d \Delta P}{d(vt)}\Bigg|_{v=0},
\\
{\rm Re}\, \sigma^W_h &\simeq \frac{d\Delta Q}{d(vt)}\Bigg|_{v=0}
-
\frac{{\rm Cov}\left(P,\,Q\right)_i}{{\rm Var}\left(P\right)_i}
\frac{d \Delta P}{d(vt)}\Bigg|_{v=0},
\end{align}
From these expressions, we learn that the weak value is characterized as the susceptibility of the probe system observables which affects the shift in the probe system.
In Section~\ref{sec:Ramsey_ind}, we generalize the Ramsey resonance to the context of indirect measurement as an example of the commutative case.

\subsubsection{Non-commutative case
}
For non-commutative case $\vec{n}^{(h)} \cdot \vec{n}^{(v)} = 0$ for which $[H_0,\, V] \neq 0$ holds, the time evolution described by Eq.~\eqref{eq:time3} becomes
\begin{align}
	{|\Lambda_f\rangle}=e^{-i H_{\rm ind} t} {|\psi_i\otimes \Psi_i\rangle}\simeq 
	\left[
	e^{-i H_0 t}\otimes I_{\mathcal{P}}
	-i \frac{v}{2h}[\sigma_a,\, e^{-i H_0 t}]\otimes P
	\right]{|\psi_i\otimes \Psi_i\rangle}.
\end{align}
The probe $\mathcal{P}$ state after the time evolution with the post-selection is then
\begin{align}
	{|\Psi_f\rangle}&\simeq {\langle \psi_f|e^{-i H_0 t}|\psi_i\rangle}
    \left[
    I_{\mathcal{P}}-i \frac{v}{2h} \left(
    \sigma^W_{a,L}-\sigma^W_{a,R}
    \right)P
    \right]{|\Psi_i\rangle}
    .\label{eq:Psif_noncom}
\end{align}
From Eq.~\eqref{eq:Psif_noncom}, we find the expectation value of $F$, 
\begin{align}
    {\langle \Psi_f|F|\Psi_f\rangle}
    &\simeq \left|{\langle \psi_f|e^{-i H_0 t}|\psi_i\rangle}\right|^2
    \notag
    \\
    &\times \bigg[
    {\langle \Psi_i|F|\Psi_i\rangle}+\frac{v}{2h} \left\{
    {\rm Re}\,\left(
    \sigma^W_{a,L}-\sigma^W_{a,R}
    \right) {\langle \Psi_i|C_{FP}|\Psi_i\rangle}
    +
    {\rm Im}\,\left(
    \sigma^W_{a,L}-\sigma^W_{a,R}
    \right) {\langle \Psi_i|A_{FP}|\Psi_i\rangle}
    \right\}
    \bigg].\label{eq:F_non}
\end{align}
%
%
The probe shift is then found to be
\begin{align}
    \Delta F&={\langle F\rangle}_f-{\langle F\rangle}_i\notag
    \\
    &\simeq \frac{v}{2h}
    \bigg[
    {\rm Re}\, \left(
    \sigma^W_{a,L}-\sigma^W_{a,R}
    \right) {\langle C_{FP}\rangle}_i
    +
    2{\rm Im}\, \left(
    \sigma^W_{a,L}-\sigma^W_{a,R}
    \right) {\rm Cov}\left(F,\,P\right)_i
    \bigg].\label{eq:shift_F_non}
\end{align}
For $F=P$ and $Q$, this yields
\begin{align}
    \Delta P&\simeq \frac{v}{h}\,
    {\rm Im}\,\left(
    \sigma^W_{a,L}-\sigma^W_{a,R}
    \right) {\rm Var}\left(P\right)_i,\label{eq:P_exp_un}
    \\
    \Delta Q&\simeq \frac{v}{2h}
    \bigg[
    {\rm Re}\, \left(
    \sigma^W_{a,L}-\sigma^W_{a,R}
    \right) 
    +
    2{\rm Im}\, \left(
    \sigma^W_{a,L}-\sigma^W_{a,R}
    \right)
    {\rm Cov}\left(P,\,Q\right)_i
    \bigg].\label{eq:X_exp_un}
\end{align}
Upon combining Eqs.~\eqref{eq:P_exp_un} and \eqref{eq:X_exp_un}, the real and imaginary parts of the difference in the weak value $\sigma^W_{a,L}-\sigma^W_{a,R}$ are expressed, respectively, as
\begin{align}
     {\rm Im}\,\left(
    \sigma^W_{a,L}-\sigma^W_{a,R}
    \right)&\simeq \frac{1}{2{\rm Var}\left(P\right)_i}\frac{d \Delta P}{d(v/2h)}\bigg|_{v=0},
    \\
    {\rm Re}\,\left(
    \sigma^W_{a,L}-\sigma^W_{a,R}
    \right)&\simeq \frac{d\Delta Q}{d(v/2h)}\bigg|_{v=0}-\frac{{\rm Cov}\left(P,\,Q\right)_i}{{\rm Var}\left(P\right)_i}\frac{d \Delta P}{d(v/2h)}\bigg|_{v=0}.
\end{align}
As in the commutative case, the weak value can be interpreted as the susceptibility of the probe system observables, and it can be extracted from the shift in the probe system.
In Section~\ref{sec:Rabi_ind}, we shall present the non-commutative case by the example of the Rabi resonance formulated in the framework of indirect measurement.

\section{Quantum resonance}
\label{sec:reso}
Before discussing the connection between the quantum resonance and the characteristic amplification seen in weak measurement, we briefly review the concept of quantum resonance with two examples: the Rabi~\cite{Rabi:1937dgo} and the Ramsey resonance~\cite{Ramsey:1950tr}.

\subsection{Rabi resonance}
Consider a qubit system governed by the following time-dependent Hamiltonian,
\begin{align}
H_{\rm Rabi}(t):=-\frac{\omega_0}{2}\sigma_3+\omega_1 \left(\cos\omega t\, \sigma_1-\sin\omega t\, \sigma_2\right),
\label{eq:hamRabi}
\end{align}
with real constants $\omega_0, \omega_1$ and $\omega$.
This describes a particle spin system subjected to an external magnetic field.
Here, $\omega_0$ represents the product of the static magnetic field along the $z$-axis and the particle's magnetic moment.
Additionally, the spin is affected by a rotating magnetic field in the $x$-$y$ plane with angular frequency $\omega$.
The symbol $\omega_1$ denotes the magnitude of the product of the rotating field and the particle's magnetic moment.
Note that the particle is also characterized by its position state, which implicitly behaves as the state of the probe.
As will be discussed in Section~\ref{sec:res_ind}, the (spin's) position-dependent magnetic fields can serve as the interaction term between the qubit and probe systems.

Now, the resonance appearing in this setting is referred to as {\it Rabi resonance} (or {\it magnetic resonance}).
The time evolution of the state ${|\psi(t)\rangle}$ is then governed by the Schr\"{o}dinger equation,
\begin{align}
    i\frac{d}{dt}{|\psi(t)\rangle}=H_{\rm Rabi}(t){|\psi(t)\rangle}.\label{eq:schIII}
\end{align}
In a rotating coordinate frame ${|\psi'(t)\rangle}:=e^{-i\omega t\sigma_3/2}{|\psi(t)\rangle}$, Eq.~\eqref{eq:schIII} is rewritten by,
$
    i{d}{|\psi'(t)\rangle}/{dt}=H'_{\rm Rabi}{|\psi'(t)\rangle}
$
with the time-independent operator,
\begin{align}
    H'_{\rm Rabi}:=\omega_1 \left(\sigma_1+\frac{\omega-\omega_0}{2\omega_1}\sigma_3 \right).\label{eq:HamRabi}
\end{align}
The time evolution of the state between $t_0$ and $t_0+t$ then reads
\begin{align}
    {|\psi(t_0+t)\rangle}=e^{i\omega (t_0+t)\sigma_3/2}e^{-iH'_{\rm Rabi} t}e^{-i\omega t_0\sigma_3/2}{|\psi(t_0)\rangle},\label{eq:timeIII}
\end{align}
where $H'_{\rm Rabi}$ can be regarded as the Hamiltonian $H$ of Eq.~\eqref{eq:Ham1}.

Now, let us choose ${|\psi(t_0)\rangle}={|\pm\rangle}$ for our initial state, where $\sigma_3 {|\pm\rangle} =\pm {|\pm\rangle}$.
From Eq.~\eqref{eq:timeIII}, we have the transition amplitude from ${|\pm\rangle}$ to ${|\pm\rangle}$,
\begin{align}
{\langle \pm| \psi(t_0+t)\rangle}=e^{\pm i\omega t/2} {\langle \pm| e^{-iH'_{\rm Rabi}t}|\pm\rangle},\label{eq:ampIII}
\end{align}
and the transition probability,
\begin{align}
    {\rm Pr}_{\pm\to\pm}^{\rm Rabi}=\left|{\langle \pm| e^{-iH'_{\rm Rabi}t}|\pm\rangle}\right|^2.\label{eq:proIII}
\end{align}
Resonance arises at $\omega=\omega_0$, {\it i.e.,} $H'_{\rm Rabi}=\omega_1\sigma_1$, with the probability,
\begin{align}
    {\rm Pr}_{\pm\to\pm}^{\rm Rabi}=\left|{\langle \pm| e^{-i\omega_1t\sigma_1}|\pm\rangle}\right|^2~{\rm for}~\omega=\omega_0.\label{eq:proIII2}
\end{align}
Specifically, for $\omega_1 t \simeq \pi/2$, the amplitude in Eq.~\eqref{eq:ampIII} is strongly suppressed, causing the probability in Eq.~\eqref{eq:proIII2} to vanish.
Conversely, under the same condition, the transition probability from ${|\pm\rangle}$ to ${|\mp\rangle}$ is strongly enhanced and approaches one, since
${\rm Pr}_{\pm \to \mp}^{\rm Rabi} = 1 - {\rm Pr}_{\pm \to \pm}^{\rm Rabi}$.
The transition probability is maximally enhanced at the resonance point $\omega=\omega_0$ for $\omega_1 t = \pi/2$, corresponding to the standard resonance case (because the significance of the resonance relies on the strong enhancement of the transition probability).
Although variations around $\omega_1 t = \pi/2$ have been studied earlier in Ref.~\cite{Ueda:2023xoj}, we focus on this typical case below, since our basic observations remain the same even for cases where
$1/2 \ll {\rm Pr}_{\pm \to \mp}^{\rm Rabi}$, when the peak transition probability is significantly larger than the complementary probability ${\rm Pr}_{\pm \to \pm}^{\rm Rabi} = 1 - {\rm Pr}_{\pm \to \mp}^{\rm Rabi}$.

The characteristic aspect of this resonance is that the probability ${\rm Pr}_{\pm \to \mp}^{\rm Rabi}$ is greatly amplified at $\omega = \omega_0$ for $\omega_1 t\simeq \pi/2$.
To see this, let us consider an experiment in which we have three adjustable parameters $\omega$, $\omega_1$, and $t$, and one unknown parameter $\omega_0$.
In order to find the unknown parameter $\omega_0$, one may vary $\omega$ while keeping $\omega_1$ and $t$ constant, and observe the peak position in the probability distribution. 
The peak then indicates the value $\omega = \omega_0$, with the accuracy determined by the peak width.
As will be explained in Section~\ref{sec:reso_dir}, this measurement process can be regarded as a weak measurement because it provides information about the slight disturbance of the system near the resonance point $\omega = \omega_0$.
In addition, we present in Section~\ref{sec:res_ind} an example of the Rabi resonance in a generalized scheme of indirect measurement.

\subsection{Ramsey resonance}
We next examine the influence of the Hamiltonian on the spin of a particle divided into the three time intervals given by
\begin{align}
    H_{\rm Ramsey}(t):=\begin{cases}
    -\frac{\omega}{2}\sigma_3+\omega_1 \left(\cos\omega t \sigma_1-\sin\omega t\sigma_2\right),~~&t_0\leq t < t_0+\tau/2,
    \\
    -\frac{\omega_0}{2}\sigma_3,~~&t_0+\tau/2\leq t < t_0+\tau/2+T,
    \\
    -\frac{\omega}{2}\sigma_3+\omega_1\left(\cos\omega t \sigma_1-\sin\omega t\sigma_2\right),~~&t_0+\tau/2+T\leq t < t_0+\tau+T.
    \end{cases}
    \label{eq:HRamsey1}
\end{align}
Since the forms of the Hamiltonian for the first and third time intervals are the same as $H_{\rm Rabi}$ in Eq.~\eqref{eq:hamRabi} with $\omega_0 = \omega$, we know that the time evolution of the initial state $|\psi(t_0)\rangle$ given by Eq.~\eqref{eq:timeIII} is
%
%
\begin{align}
    {|\psi(t_0+\tau+T)\rangle}=e^{i\omega (t_0+\tau+T)\sigma_3/2}e^{-i\omega_1 \tau \sigma_1/2}e^{-i (\omega-\omega_0) T\sigma_3/2} e^{-i \omega_1\tau\sigma_1/2}e^{-i \omega t_0\sigma_3/2}{|\psi(t_0)\rangle}.\label{eq:timramsey}
\end{align}
%
%
As in the case of the Rabi resonance, we choose the initial state ${|\psi(t_0)\rangle} = {|\pm\rangle}$.
According to Eq.~\eqref{eq:timramsey}, the transition amplitude from the state ${|\pm\rangle}$ to ${|\pm\rangle}$ is given by
\begin{align}
    {\langle \pm|\psi(t_0+\tau+T)\rangle}&=e^{\pm i\omega (\tau+T)/2} {\langle \pm|}e^{-i\omega_1 \tau \sigma_1/2}e^{-i(\omega-\omega_0)T \sigma_3/2}e^{-i\omega_1\tau \sigma_1/2}{|\pm\rangle}.\label{eq:ampRamsey}
\end{align}
From Eq.~\eqref{eq:ampRamsey}, we find the transition probability,
\begin{align}
    {\rm Pr}_{\pm\to\pm}^{\rm Ramsey}=\left|{\langle \pm|}e^{-i\omega_1 \tau \sigma_1/2}e^{-i(\omega-\omega_0)T \sigma_3/2}e^{-i\omega_1\tau \sigma_1/2}{|\pm\rangle}\right|^2.\label{eq:prRamsey}
\end{align}
As before, we further analyze the case $\omega=\omega_0$ to observe
\begin{align}
    {\rm Pr}_{\pm\to\pm}^{\rm Ramsey}=\left|{\langle \pm|}e^{-i\omega_1 \tau \sigma_1}{|\pm\rangle}\right|^2~{\rm for}~\omega=\omega_0.~\label{eq:proIIIramsey}
\end{align}
Notice that Eq.~\eqref{eq:proIIIramsey} has the same form as Eq.~\eqref{eq:proIII2}, and at $\omega_1\tau=\pi/2$ the probability vanishes. This emergence of resonance is known as {\it Ramsey resonance}.
%
%
At this point, we recognize the fundamental difference between the Ramsey resonance and the Rabi resonance lies in the fact that,
while the Rabi resonance involves two oscillations $\omega$ and $\omega_0$ occurring within the same time interval, the Ramsey resonance is characterized by oscillations occurring in separate time intervals.
%

\section{Quantum resonance viewed as weak measurement}
\label{sec:weak_reso}
To clarify the link of the quantum resonance explained in Section~\ref{sec:reso} to the weak measurement, we first reconsider the quantum resonance in light of the direct weak measurement, 
and then expand our argument of the link to the indirect weak measurement.

\subsection{Quantum resonance in the direct measurement}
\label{sec:reso_dir}
We now reconsider the two typical quantum resonances discussed in Section~\ref{sec:reso} without explicitly introducing a probe system.
%
%
In the following, we interpret a small deviation from the resonance condition as a source of interaction associated with a weak measurement process.  This allows us to regard the resonance as the amplification characteristic to the weak measurement.

\subsubsection{Rabi resonance}
\label{sec:Rabi_dir}
Let us first revisit the Rabi resonance phenomena in the framework of direct weak measurement.
To clarify the connection between the weak measurement and the resonance, we start by putting the parameter $\omega_0$ as
\begin{align}
    \omega_0:=\overline{\omega}_0+\epsilon,
    \label{eq:ep1}
\end{align}
and regard $\epsilon$ as a parameter representing the small disturbance associated with the measurement.  
More explicitly, our weak measurement via Rabi resonance requires the following three assumptions:
(i) $\overline{\omega}_0$, $\omega$, $\omega_1$, and $t$ are known parameters;
(ii) $\epsilon$ is an unknown parameter; and
(iii) the condition $\epsilon/\omega_1 \ll 1$ characterizes a weak measurement.
We remark that the last condition $\epsilon/\omega_1 \ll 1$ indicates a small perturbation from the null case $\epsilon = 0$, which is a necessary ingredient of the weak measurement.  
With these assumptions in mind, we write the Hamiltonian in Eq.~\eqref{eq:HamRabi} as
\begin{align}
    H'_{\rm Rabi}=\omega_1 \sigma_1+\omega_1\left(\phi_{\rm Rabi}+\delta\right)\sigma_3,
    \label{eq:Rabipri}
\end{align}
with
\begin{align}
\phi_{\rm Rabi}:=\frac{\omega-\overline{\omega}_0}{2\omega_1}
   \label{eq:Rabiphase}
\end{align}
and $\delta:= -\epsilon/2\omega_1$.  
Although Eq.~\eqref{eq:Rabipri} shows that the disturbance $\epsilon$ can be absorbed into $\phi_{\rm Rabi}$, we retain the parameter $\epsilon$ in order to clarify its connection with the weak measurement.

To accommodate the expression $H'_{\rm Rabi} = H$ in Eq.~\eqref{eq:Ham1}, we define the two operators $H_0$ and $V$ by 
\begin{align}
    H_0&=\omega_1\left( \sigma_1+\phi_{\rm Rabi}\, \sigma_3\right),\qquad V=\omega_1\delta\,\sigma_3.\label{eq:HamRab}
\end{align}
In what follows, we concentrate on a parameter region near the resonance point satisfying $\phi_{\rm Rabi} \ll 1$ and $\delta\ll 1$.
The coefficients of the Pauli matrices of Eq.~\eqref{eq:h0V} are determined, up to the first order of $\phi_{\rm Rabi}$ and $\delta$, as
\begin{align}
    &h=\omega_1,\qquad n_0^{(h)}=0,\qquad \vec{n}^{(h)}= \left(1,\,0,\,\phi_{\rm Rabi}\right),\notag
    \\
    &v=\omega_1\delta,\qquad n_0^{(v)}=0,\qquad \vec{n}^{(v)}=\left(0,\,0,\,1\right).
\end{align}
Using $\sigma_h = \sigma_1+\phi_{\rm Rabi}\, \sigma_3$ and $\sigma_v = \sigma_3$, and  
taking $\sigma_a=-\sigma_2$, we find 
\begin{align}
[\sigma_a,\, \sigma_h]=2i (\sigma_3 + \phi_{\rm Rabi}\, \sigma_1),
\end{align}
which corresponds to the condition \eqref{eq:condsa}.  
From Eq.~\eqref{eq:time2}, the time evolution of the system gives
\begin{align}
    &{|\psi(t_0+t)\rangle}\simeq  \left(
    e^{-iH_0t}+i\frac{\delta}{2} \left[\sigma_2,\,
    e^{-iH_0t}\right]\right){|\psi_i\rangle}\notag
    \\
    &\simeq  \left(e^{i\phi_{\rm Rabi}\, \sigma_2/2}
    e^{-i\omega_1t\sigma_1}e^{-i\phi_{\rm Rabi}\, \sigma_2/2}+i\frac{\delta}{2} \left[\sigma_2,\,e^{i\phi_{\rm Rabi}\, \sigma_2/2}
    e^{-i\omega_1t\sigma_1}e^{-i\phi_{\rm Rabi}\, \sigma_2/2}\right]\right){|\psi_i\rangle},\label{eq:RabiTIM}
\end{align}
where we have used $e^{i\phi_{\rm Rabi}\,\sigma_2/2}\sigma_1 e^{-i\phi_{\rm Rabi}\,\sigma_2/2}=\sigma_1\cos\phi_{\rm Rabi}+\sigma_3\sin\phi_{\rm Rabi}$.
Upon substituting Eq.~\eqref{eq:RabiTIM} into Eq.~\eqref{eq:amp1}, we find the transition amplitude,
\begin{align}
    {\langle \psi_f|\psi(t_0+t)\rangle}\simeq {\langle \psi_f(\phi_{\rm Rabi}) |
    e^{-i\omega_1 t \sigma_1}|\psi_i(\phi_{\rm Rabi})\rangle}\left(1+i\frac{\delta}{2}\left(\sigma^W_{2,L}\left(\phi_{\rm Rabi}\right)-\sigma^W_{2,R}\left(\phi_{\rm Rabi}\right)\right)\right),\label{eq:RABAMP}
\end{align}
where ${|\psi_{i}(\phi_{\rm Rabi})\rangle}:=e^{-i \phi_{\rm Rabi} \sigma_2/2}{|\psi_i\rangle}$, ${|\psi_{f}(\phi_{\rm Rabi})\rangle}:=e^{-i \phi_{\rm Rabi} \sigma_2/2}{|\psi_f\rangle}$, with the weak values,
\begin{align}
\sigma^W_{2,L}\left(\phi_{\rm Rabi}\right)&:=\frac{{\langle \psi_f(\phi_{\rm Rabi})|\sigma_2 e^{-i\omega_1t \sigma_1}|\psi_i(\phi_{\rm Rabi})\rangle}}{{\langle \psi_f(\phi_{\rm Rabi})|e^{-i\omega_1t \sigma_1}|\psi_i(\phi_{\rm Rabi})\rangle}},
\\
\sigma^W_{2,R}\left(\phi_{\rm Rabi}\right)&:=\frac{{\langle \psi_f(\phi_{\rm Rabi})| e^{-i\omega_1t \sigma_1}\sigma_2|\psi_i(\phi_{\rm Rabi})\rangle}}{{\langle \psi_f(\phi_{\rm Rabi})| e^{-i\omega_1t \sigma_1}|\psi_i(\phi_{\rm Rabi})\rangle}}.\label{eq:weakR}
\end{align}

Now, choosing ${|\psi_i\rangle}={|\pm\rangle}$, ${|\psi_f\rangle}={|\pm\rangle}$ and plugging Eq.~\eqref{eq:RABAMP} into Eq.~\eqref{eq:pr1}, we obtain the transition probability,
\begin{align}
    {\rm Pr}_{\pm\to \pm}^{\rm Rabi}(\epsilon)\simeq {\rm Pr}_{\pm\to\pm}^{\rm Rabi}(0)  \left(1+\frac{\epsilon}{2\omega_1}\left({\rm Im}\,\sigma_{2,L}^W\left(\phi_{\rm Rabi}\right)-{\rm Im}\,\sigma_{2,R}^W\left(\phi_{\rm Rabi}\right)\right)\right).
   \label{eq:propm}
\end{align}
In particular, for $\omega_1t=\pi/2$ (see also Ref.~\cite{Ueda:2023xoj} for outcomes for deviations from $\omega_1t=\pi/2$), we obtain the weak values,
\begin{align}
    \sigma^W_{2,L}\left(\phi_{\rm Rabi}\right)=-\sigma^W_{2,R}\left(\phi_{\rm Rabi}\right)=-i\cot\phi_{\rm Rabi}.\label{eq:weaksRab}
\end{align}
With these, the probability for $\epsilon =0$ reads
\begin{align}
    {\rm Pr}_{\pm \to\pm}^{\rm Rabi}(0)&=\left|{\langle \psi_f(\phi_{\rm Rabi})|e^{-i\omega_1t \sigma_1}|\psi_i(\phi_{\rm Rabi})\rangle}\right|^2=\sin^2 \phi_{\rm Rabi}.
\end{align}
The transition probability Eq.~\eqref{eq:propm} is then found to be
\begin{align}
    {\rm Pr}_{\pm\to\pm}^{\rm Rabi}(\epsilon)\simeq {\rm Pr}_{\pm\to\pm}^{\rm Rabi}(0)  \left(1+\delta_{\rm Rabi}\, {\rm Im}\,\sigma_{2,L}^W\left(\phi_{\rm Rabi}\right)\right),\label{eq:Prr}
\end{align}
where $\delta_{\rm Rabi}:=2\epsilon t/\pi$ denotes the measurement strength of the Rabi resonance for $\omega_1t=\pi/2$.
Note that at the point $\phi_{\rm Rabi}=0$, the probability \eqref{eq:Prr} vanishes, ${\rm Pr}_{\pm\to \pm}^{\rm Rabi}(\epsilon)=0$, indicating the appearance of resonance, and at this precise resonance point the weak values \eqref{eq:weaksRab} are simultaneously amplified.
This shows that Rabi resonance can be seen as weak value amplification with measurement strength $\delta_{\rm Rabi}$, and that the Rabi resonance condition is nothing but the divergence condition of the weak value.
It is then obvious that the Rabi resonance can be seen as a weak value amplification even for values of $\omega_1 t\neq \pi/2$ that satisfy $1/2 \ll {\rm Pr}_{\pm\to\mp}^{\rm Rabi}$, with the peak significantly enhanced~\cite{Ueda:2023xoj}.

We also note that the imaginary part of the weak value of $\sigma_2$ can be derived from Eq.~\eqref{eq:Prr} as
\begin{align}
    {\rm Im}\,\sigma^W_{2,L}(\phi_{\rm Rabi})=\frac{1}{{\rm Pr}^{\rm Rabi}_{\pm\to\pm}(0)}\frac{d{\rm Pr}^{\rm Rabi}_{\pm\to\pm}(\epsilon)}{d\delta_{\rm Rabi}}\bigg|_{\delta_{\rm Rabi}=0}.\label{eq:weakrab}
\end{align}
%
%
This shows that the measurement of $\epsilon$ utilizing the Rabi resonance  extracts in effect the imaginary component of the weak value of $\sigma_2$.
%
%

%

\subsubsection{Ramsey resonance}
\label{sec:Ramsey_direct}
Next, we examine the Ramsey resonance from the perspective of weak measurement.
As we did in Eq.~\eqref{eq:ep1}, we first express the parameter $\omega_0$ as $\omega_0 := \overline{\omega}_0 + \epsilon$, and
consider the weak measurement of $\epsilon$ under the following assumptions:
(i) $\overline{\omega}_0$, $\omega$, $\omega_1$, $\tau$, and $T$ are known parameters;
(ii) $\epsilon$ is an unknown parameter; and
(iii) the condition for weak measurement is satisfied when $\epsilon T \ll 1$.
During the time interval $ t_0 + \frac{\tau}{2} \leq t < t_0 + \frac{\tau}{2} + T $, the Hamiltonian in Eq.~\eqref{eq:HRamsey1} can be rewritten as
\begin{align}
    H_{\rm Ramsey}(t)=-\frac{\overline{\omega}_0}{2}\sigma_3-\frac{\epsilon}{2}\sigma_3.\label{eq:ramsec}
\end{align}
Based on the above setup, we choose the two operators $H_0$ and $V$ in \eqref{eq:Ham1} by
\begin{align}
    H_0 =-\frac{\overline{\omega}_0}{2}\sigma_3,\qquad V=-\frac{\epsilon}{2} \sigma_3,
\end{align}
and the coefficients in Eq.~\eqref{eq:h0V},
\begin{align}
    &h=\frac{\overline{\omega}_0}{2},\qquad n^{(h)}_0=0,\qquad \vec{n}^{(h)}=\left(0,\,0,\,-1\right),\notag
    \\   &v=\frac{\epsilon}{2},\qquad n_0^{(v)}=0,\qquad \vec{n}^{(v)}=\left(0,\,0,\,-1\right).
\end{align}
These implies $\vec{n}^{(h)}\times \vec{n}^{(v)}=0$, showing that the case of the Ramsey resonance belongs to the commutative case mentioned in Section~\ref{sec:direct}.

Now, we observe that our intial and final states,
\begin{align}
{|\psi_i\rangle}:&=e^{i\omega (t_0+\tau/2)\sigma_3/2}e^{-i\omega_1\tau\sigma_1/2}e^{-i\omega t_0\sigma_3/2}{|\pm\rangle},
\\
{|\psi_f\rangle}:&=e^{i\omega (t_0+\tau/2+T)\sigma_3/2}e^{i\omega_1\tau\sigma_1/2}e^{-i\omega(t_0+\tau+T)\sigma_3/2}{|\pm\rangle},
\end{align}
combined with Eqs.~\eqref{eq:PrI} and \eqref{eq:ramsec}, yield the transition probability from ${|\psi_i\rangle}$ to ${|\psi_f\rangle}$,
\begin{align}
    {\rm Pr}_{\pm\to \pm}^{\rm Ramsey}(\epsilon)\simeq {\rm Pr}_{\pm\to \pm}^{\rm Ramsey}(0) \left(1- \epsilon T\,{\rm Im}\, \sigma_3^W\right),
\end{align}
where
\begin{align}
    {\rm Pr}_{\pm\to \pm}^{\rm Ramsey}(\epsilon):=\left|{\langle \psi_f| e^{-i\int_{t_0+\tau/2}^{t_0+\tau/2+T}dt H_{\rm Ramsey}(t)}|\psi_i\rangle} \right|^2,\label{eq:PrRamseyif}
\end{align}
and
\begin{align}
    \sigma_3^W =\frac{{\langle \psi_f|e^{-i H_0 T}\sigma_3|\psi_i\rangle}}{{\langle \psi_f| e^{-i H_0 T}|\psi_i\rangle}}=\frac{{\langle \psi_f|\sigma_3e^{-i H_0 T}|\psi_i\rangle}}{{\langle \psi_f| e^{-i H_0 T}|\psi_i\rangle}}.\label{eq:weak3}
\end{align}
In particular, at $\omega_1\tau = \pi/2$ we have the relations,
\begin{align}
    &{\langle \psi_f|e^{-i H_0 T }\sigma_3|\psi_i\rangle}=e^{\pm i\omega(\tau+T)/2}{\langle \psi_f(\phi_{\rm Ramsey})| e^{-i\omega_1\tau \sigma_1}\sigma_2 |\psi_i(\phi_{\rm Ramsey})\rangle},\label{eq:we1p}
\\
&{\langle \psi_f|e^{-i H_0 T }|\psi_i\rangle}=e^{\pm i\omega(\tau+T)/2}{\langle \psi_f(\phi_{\rm Ramsey})|e^{-i\omega_1\tau \sigma_1} |\psi_i(\phi_{\rm Ramsey})\rangle},\label{eq:we2p}
\end{align}
where
\begin{align}
{|\psi_i(\phi_{\rm Ramsey})\rangle}={|\psi_f(\phi_{\rm Ramsey})\rangle}:=e^{-i\phi_{\rm Ramsey}\, \sigma_2/2}{|\pm\rangle},
\end{align}
with 
\begin{align}
\phi_{\rm Ramsey}:=(\omega-\overline{\omega}_0)\frac{T}{2}.
   \label{eq:Ramseyphase}
\end{align}
As in the case of the Rabi resonance discussed before, although the disturbance parameter $\epsilon$ can be absorbed into $\phi_{\rm Ramsey}$ as seen from Eqs.~\eqref{eq:ramsec} and \eqref{eq:Ramseyphase}, we retain it to clarify the connection between the resonance and the weak measurement.

Let us now notice that, with Eqs.~\eqref{eq:we1p} and \eqref{eq:we2p}, we have
\begin{align}
    \sigma_3^W =- \sigma_{2,L}^W\left(\phi_{\rm Ramsey}\right)= \sigma_{2,R}^W\left(\phi_{\rm Ramsey}\right)= i\cot \phi_{\rm Ramsey},\label{eq:weak3Ramsey}
\end{align}
and also for $\epsilon=0$,
\begin{align}
    {\rm Pr}_{\pm\to \pm}^{\rm Ramsey}(0)&=\left|{\langle \psi_f(\phi_{\rm Ramsey})|e^{-i\omega_1\tau \sigma_1} |\psi_i(\phi_{\rm Ramsey})\rangle}\right|^2=\sin^2\phi_{\rm Ramsey}.
\end{align}
From these, we find that Eq.~\eqref{eq:PrRamseyif} reduces to
\begin{align}
    {\rm Pr}_{\pm\to \pm}^{\rm Ramsey}(\epsilon)\simeq {\rm Pr}_{\pm\to \pm}^{\rm Ramsey}(0) \left(1+\delta_{\rm Ramsey}\,{\rm Im}\,\sigma^W_{2,L}\left(\phi_{\rm Ramsey}\right) \right), \label{eq:proramseyf}
\end{align}
where the measurement strength for the Ramsey resonance is defined as $\delta_{\rm Ramsey}:=\epsilon T$, and the imaginary part of $\sigma^W_{2,L}$ is provided in Eq.~\eqref{eq:weaksRab}.
As we have seen in the Rabi resonance, for $\phi_{\rm Ramsey}=0$ the probability described by Eq.~\eqref{eq:proramseyf} approaches zero, leading to a divergence in the weak value as noted in Eq.~\eqref{eq:weak3Ramsey}.
This suggests that the Ramsey resonance can also be considered a weak value amplification characterized by the measurement strength $\delta_{\rm Ramsey}$.

It is noteworthy that, aside from the difference in measurement strength, the two probability formulae, Eqs.~\eqref{eq:Prr} and \eqref{eq:proramseyf}, become identical when the condition $T = t$ is satisfied.
In particular, we find that $\phi_{\rm Ramsey} = \phi_{\rm Rabi}$ and $\delta_{\rm Ramsey} = \delta_{\rm Rabi}$ when $T$ is substituted with $\frac{2t}{\pi}$.
The weak measurement framework thus offers a unified approach to understanding the two types of resonances with fundamentally different characteristics as mentioned above, where the difference can also be seen from the fact that the two are classified with respect to the commutative and non-commutative types as discussed in Section~\ref{sec:weak_qubit}. 

Another point to be noted is that in the weak measurement framework the measurement strength plays a crucial role in quantifying the resonance's sensitivity to small variations in resonance conditions.
%
%
%
In this respect, it is important to see from Eq.~\eqref{eq:proramseyf} that the imaginary component of the weak value admits the form,
\begin{align}
    {\rm Im}\,\sigma^W_{2,L}(\phi_{\rm Ramsey})=\frac{1}{{\rm Pr}_{\pm\to\pm}^{\rm Ramsey}(0)}
    \frac{d {\rm Pr}_{\pm\to\pm}^{\rm Ramsey}(\epsilon)}{d\delta_{\rm Ramsey}}\bigg|_{\delta_{\rm Ramsey}=0}.\label{eq:weakRams0}
\end{align}
Thus, as in the case of the Rabi resonance, we see that the Ramsey resonance also determines the imaginary component of the weak value of $\sigma_2$ as an indicator of the resonance’s sensitivity.
%

\subsection{Quantum resonance in the indirect measurement}
\label{sec:res_ind}
As seen in Section~\ref{sec:reso_dir}, both the Rabi resonance and the Ramsey resonance can be interpreted as a direct weak measurement.  Here we show that both of the two resonances can also be put into a unified framework of indirect weak measurement where the probe system is explicitly involved.

\subsubsection{Rabi resonance}
\label{sec:Rabi_ind}
We begin by the case of the Rabi resonance realized as an indirect weak measurement.
Recall the qubit spin system of a particle introduced in Section~\ref{sec:reso_dir}.
Since a particle carries freedom of position besides the spin, the total space of quantum states consists of the space corresponding the position $\mathcal{P}$ in addition to the spin $\mathcal{S}$.  Here we regard $\mathcal{P}$ as a probe system, and
for simplicity we restrict our attention to two distinct path states of the particle denoted ${|\mathrm{I}\rangle}$ and ${|\mathrm{II}\rangle}$, implying that our space $\mathcal{P}$ becomes effectively the path space spanned by the two states.  
Such a setup can be implemented using a beam splitter decomposing the particle's trajectories into the two paths.

We then consider the time evolution which differs depending on the paths prescribed by the Hamiltonian:
\begin{align}
	H_{\rm Rabi}^{(\rm Ind)}(t) := H_{\rm Rabi}^{(\rm I)}(t)\otimes {|\rm I\rangle} {\langle \rm I|}+ H_{\rm Rabi}^{(\rm II)}(t)\otimes {|\rm II\rangle} {\langle \rm II|},\label{eq:Ham_Ind_Rabi}
\end{align}
with
\begin{align}
H^{\rm (I)}_{\rm Rabi}(t):&= 
	-\frac{\omega_0}{2}\sigma_3 +\omega_1 \left(\cos\omega t \sigma_1 -\sin \omega t \sigma_2\right),\label{eq:Rabi_I}
	\\
H^{\rm (II)}_{\rm Rabi}(t):&= +\frac{\omega_0}{2}\sigma_3 +\omega_1 \left(\cos\omega t \sigma_1 +\sin \omega t \sigma_2\right).\label{eq:Rabi_II}
\end{align}
Since the introduction of $\epsilon$ made previously does not affect the physics, as noted before, we shall temporarily omit it.
The Hamiltonians for the qubit system, Eqs.~\eqref{eq:Rabi_I} and \eqref{eq:Rabi_II}, are formally identical to Eq.~\eqref{eq:hamRabi}, but the external fields differ depending on the path states ${|\mathrm{I}\rangle}$ and ${|\mathrm{II}\rangle}$.
As evident from the Hamiltonians, the Rabi resonance can occur independently along each path.
Note, however, that when we enforce the same time evolution for the particles on the two paths, $H^{\rm (I)}_{\rm Rabi}(t) = H^{\rm (II)}_{\rm Rabi}(t)$, then obviously the path dependence becomes unobservable.
In that event, the system reduces effectively to the direct measurement described by Eq.~\eqref{eq:tevco}, as long as we restrict ourselves to the subspace spanned by ${|\mathrm{I}\rangle}$ and ${|\mathrm{II}\rangle}$. 
This is because the total Hamiltonian in the indirect setting, $H^{(\mathrm{Ind})}_{\rm Rabi}(t)$, becomes proportional to ${|\mathrm{I}\rangle}{\langle \mathrm{I}|} + {|\mathrm{II}\rangle}{\langle \mathrm{II}|} = I_{\mathcal{P}}$ which is the identity operator in the path space.  
This suggests that, in the context of the Rabi resonance, the direct weak measurement arises as a special case of the indirect weak measurement.

Now, we follow the above setup and consider the time evolution of this system.
According to Eq.~\eqref{eq:Ham_Ind_Rabi}, the time evolution operator in this system reads
\begin{align}
	e^{-i \int_{t_0}^{t_0+t} dt H_{\rm Rabi}^{(\rm Ind)}(t) }
	&=e^{i \omega (t_0 +t)\sigma_3/2} e^{-i H'_{\rm Rabi}(\omega,\, \omega_0,\, \omega_1) t} e^{-i \omega t_0 \sigma_3/2}
	\otimes {|\rm I\rangle}{\langle \rm I|} \notag
	\\
	&\qquad+
e^{-i \omega (t_0 +t)\sigma_3/2} e^{-i H'_{\rm Rabi}(-\omega,\, -\omega_0,\, \omega_1) t} e^{+i \omega t_0 \sigma_3/2}
	\otimes {|\rm II\rangle}{\langle \rm II|}, 
\end{align}
where $H'_{\rm Rabi}(\omega,\, \omega_0,\, \omega_1):=H'_{\rm Rabi}$ is defined from Eq.~\eqref{eq:HamRabi}.
For an initial state of the total system ${|\Lambda_i\rangle}:={|\psi_i\otimes \Psi_i\rangle}$, where ${|\psi_i\rangle}$ is an initial state of the qubit system and ${|\Psi_i\rangle}$ is that of the path state, the time evolution of the state between $t_0$ and $t_0+t$ becomes
\begin{align}
	{|\Lambda_f\rangle} :=& \,e^{-i \int_{t_0}^{t_0+t} dt H_{\rm Rabi}^{(\rm Ind)}(t) } {|\Lambda_i\rangle}\notag
    \\
 =& \,{\langle {\rm I}|\Psi_i\rangle}\left(e^{i \omega (t_0 +t)\sigma_3/2} e^{-i H'_{\rm Rabi}(\omega,\, \omega_0,\, \omega_1) t} e^{-i \omega t_0 \sigma_3/2}\otimes I_{\mathcal{P}}\right)  {|\psi_i\otimes {\rm I}\rangle }\notag
 \\
	&\qquad+
	{\langle {\rm II}|\Psi_i\rangle} \left(e^{-i \omega (t_0 +t)\sigma_3/2} e^{-i H'_{\rm Rabi}(-\omega,\,-\omega_0,\,\omega_1) t} e^{+i \omega t_0 \sigma_3/2}\otimes I_{\mathcal{P}} \right) {|\psi_i\otimes {\rm II}\rangle }.\label{eq:Rabi_ind_time}
\end{align}
As in the direct weak measurement in Section~\ref{sec:reso_dir}, we consider ${|\psi_i\rangle}={|\pm\rangle}$ as the initial state of the qubit system and ${|\psi_f\rangle}={|\pm\rangle}$ as the post-selected state.
Then, the post-selection after the time-evolution~\eqref{eq:Rabi_ind_time} yields the path state,
\begin{align}
{|\Psi_f\rangle}:&= {\langle \psi_f|\Lambda_f\rangle}\notag
	\\
&= {\langle \pm|} {\rm exp}\bigg[
-it \bigg(H'_{\rm Rabi}(\omega,\, \omega_0,\, \omega_1)\otimes {|\rm I\rangle}{\langle \rm I|} 
+
H'_{\rm Rabi}(-\omega,\, -\omega_0,\, \omega_1)\otimes {|\rm II\rangle}{\langle \rm II|} 
\bigg) 
\bigg] {|\pm\otimes \Psi_i( \omega t)\rangle},\label{eq:Psif_rabi_ind}
\end{align}
where we have introduced ${|\Psi_i(\omega t)\rangle}:= e^{\pm i\omega t/2} {\langle {\rm I}|\Psi_i\rangle}{|{\rm I}\rangle}+e^{\mp i\omega t/2}{\langle {\rm II}|\Psi_i\rangle}{|{\rm II}\rangle}$ for brevity.
The time evolution operator appearing in Eq.~\eqref{eq:Psif_rabi_ind} becomes
\begin{align}
	&H'_{\rm Rabi}(\omega,\, \omega_0,\, \omega_1)\otimes {|\rm I\rangle}{\langle \rm I|} 
+
H'_{\rm Rabi}(-\omega,\, -\omega_0,\, \omega_1)\otimes {|\rm II\rangle}{\langle \rm II|}\notag
\\
&\quad\quad\quad\quad\quad\quad\quad\quad\quad\quad\quad\quad\quad\quad=\omega_1\bigg[ \sigma_1\otimes I_{\mathcal{P}} + \frac{\omega-\omega_0}{2\omega_1}\sigma_3 \otimes\left(
 {|\rm I\rangle} {\langle \rm I|}
 -
  {|\rm II\rangle} {\langle \rm II|}
\right)
\bigg].\label{eq:tim_rabi_ind}
\end{align}
%
%
By interpreting Eq.~\eqref{eq:tim_rabi_ind} as the Hamiltonian in Eq.~\eqref{eq:Ham2}, we find the corresponding operators in Eq.~\eqref{eq:Ham2} as 
\begin{align}
	H_0 = \omega_1\sigma_1,\qquad  V =\omega_1 \phi_{\rm Rabi} \sigma_3 
    ,\label{eq:H0_V_ind_Rabi}
\end{align}
with $\phi_{\rm Rabi}=(\omega-\omega_0)/2\omega_1$.
Also, we recognize the operator acting in the probe system in Eq.~\eqref{eq:tim_rabi_ind},
\begin{align}
	P = {|\rm I\rangle} {\langle \rm I|}
 -
  {|\rm II\rangle} {\langle \rm II|}.
  \label{eq:P}
\end{align}
We here obtain the coefficients in Eq.~\eqref{eq:h0V} as 
\begin{align}
	&h= \omega_1,\qquad n_0^{(h)}=0,\qquad  \vec{n}^{(h)} =(1,\,0,\,0),
	\\
	&v = \omega_1 \phi_{\rm Rabi},\qquad n_0^{(v)}=0,\qquad \vec{n}^{(v)}=(0,\,0,\,1).
\end{align}
From these, we can take $\sigma_a=-\sigma_2$, which fulfills the following relation, 
\begin{align}
	[\sigma_a,\,\sigma_h] =2i \sigma_3.
\end{align}
Hence, from Eq.~\eqref{eq:time3}, we find
\begin{align}
	{|\Psi_f\rangle}&\simeq {\langle \psi_f|e^{-i H_0 t}|\psi_i\rangle}\left(
	I_{\mathcal{P}}-i \frac{v}{2h}
	\frac{{\langle \psi_i|[\sigma_a,\,e^{-i H_0 t}]| \psi_f\rangle}}{{\langle \psi_f|e^{-i H_0 t}|\psi_i\rangle}} P
	\right) {|\Psi_i (\omega t)\rangle}\notag
    \\
    &={\langle \psi_f|e^{-i H_0 t}|\psi_i\rangle}
    \left[
    I_{\mathcal{P}}-i \frac{v}{2h} \left(
    \sigma^W_{a,L}-\sigma^W_{a,R}
    \right)P
    \right]{|\Psi_i (\omega t)\rangle}.
\end{align}
Recalling that the shift of the pointer represented by $F$ is given by Eq.~\eqref{eq:shift_F_non}, if we choose, 
for instance, $F=P_{\rm I}:={|\rm I\rangle}{\langle \rm I|}$ and $P_{\rm I+II}:=({|\rm I\rangle}+{|\rm II\rangle})({\langle \rm I|}+{\langle \rm II|})/2$, we obtain
\begin{align}
    \Delta P_{\rm I} &= {\langle P_{\rm I}\rangle}_f-{\langle P_{\rm I}\rangle}_i\notag
    \\
    &\simeq \frac{v}{h} {\rm Im}\, 
    \left(
    \sigma^W_{a,L}-\sigma^W_{a,R}
    \right)
    {\rm Cov}\left(P,\,P_{\rm I}\right)_i,
    \\
    \Delta P_{\rm I+II}&={\langle P_{\rm I+II}\rangle}_f-{\langle P_{\rm I+II}\rangle}_i\notag
    \\
    &\simeq \frac{v}{2h}
    \bigg[
    {\rm Re}\, \left(
    \sigma^W_{a,L}-\sigma^W_{a,R}
    \right) {\langle C_{P_{\rm I+II}P}\rangle}_i
    +
    2{\rm Im}\, \left(
    \sigma^W_{a,L}-\sigma^W_{a,R}
    \right){\rm Cov}\left(P,\,P_{\rm I+II}\right)_i
    \bigg],\label{eq:rabi_delP_I_II}
\end{align}
where $v/h=\phi_{\rm Rabi}=(\omega-\omega_0)/2\omega_1$, and ${\langle \mathcal{O}\rangle}_i := {\langle \Psi_i(\omega t)|\mathcal{O}|\Psi_i(\omega t)\rangle}/{\langle \Psi_i(\omega t)|\Psi_i(\omega t)\rangle}$ for any operator $\mathcal{O}$.
Let us introduce a small parameter $\epsilon$ through $\omega_0 = \overline{\omega}_0 + \epsilon$ to represent a perturbation
around the resonance point $\omega = \overline{\omega}_0$.
Focusing on the deviations from the resonance, we obtain $v/h = \epsilon t/\pi = \delta_{\rm Rabi}/2$ for $\omega_1 t = \pi/2$.
Thus, analogously to the direct measurement in Section~\ref{sec:Rabi_dir}, we find that $\delta_{\rm Rabi}$ characterizes the measurement strength of the Rabi resonance at $\omega_1 t = \pi/2$.

Combining these results, the weak values can be expressed in terms of the susceptibilities,
\begin{align}
    {\rm Im}\, 
    \left(
    \sigma^W_{a,L}-\sigma^W_{a,R}
    \right)&\simeq \frac{1}{2{\rm Cov}\left(P,\,P_{\rm I}\right)_i}\frac{d \Delta P_{\rm I}}{d(v/2h)}\Bigg|_{v=0},
    \\
    {\rm Re}\, 
    \left(
    \sigma^W_{a,L}-\sigma^W_{a,R}
    \right)&\simeq \frac{1}{{\langle C_{P_{\rm I+II}P}\rangle}_i} 
    \left[\frac{d \Delta P_{\rm I+II}}{d(v/2h)}\Bigg|_{v=0}
    -
    \frac{{\rm Cov}\left(P,\,P_{\rm I+II}\right)_i}{{\rm Cov}\left(P,\,P_{\rm I}\right)_i}\frac{d \Delta P_{\rm I}}{d(v/2h)}\Bigg|_{v=0}\right]
\end{align}
where
\begin{align}
    \sigma^W_{a,L}(\omega_1)&=-\frac{{\langle \pm|\sigma_2 e^{-i\omega_1 t\sigma_1}|\pm\rangle}}{{\langle \pm|e^{-i\omega_1 t\sigma_1}|\pm\rangle}}=\pm\tan \omega_1 t,
    \\
    \sigma^W_{a,R}(\omega_1)&=-\frac{{\langle \pm|e^{-i\omega_1 t\sigma_1}\sigma_2 |\pm\rangle}}{{\langle \pm|e^{-i\omega_1 t\sigma_1}|\pm\rangle}}=\mp\tan\omega_1 t.
\end{align}
Note that these weak values become maximally enhanced in the limit $\omega_1 t = \pi/2$, which corresponds to the resonance point with the maximal peak, {\it i.e.}, the point where the weak value amplification maximizes.
This implies that larger weak values can induce larger shifts in the probe system~\eqref{eq:rabi_delP_I_II}.
We also note that the above Rabi resonance allows the extraction of the real part of the weak value, in contrast to what is obtained in direct measurement, where the Hamiltonians satisfy $H^{\rm (I)}_{\rm Rabi}(t) = H^{\rm (II)}_{\rm Rabi}(t)$.

\subsubsection{Ramsey resonance}
\label{sec:Ramsey_ind}
Next we discuss the Ramsey resonance in the indirect measurement scheme.  Consider,
as before, the total system consisting of a spin $\mathcal{S}$ and a probe $\mathcal{P}$ consisting of the two path states ${|\rm I\rangle}$ and ${|\rm II\rangle}$.
With this setup, we look at the time evolution governed by the Hamiltonian,
\begin{align}
	H_{\rm Ramsey}^{(\rm Ind)}(t):=
	 H_{\rm Ramsey}^{(\rm I)}(t)\otimes {|\rm I\rangle} {\langle \rm I|}	
	+
	H_{\rm Ramsey}^{(\rm II)}(t)\otimes {|\rm II\rangle} {\langle \rm II|},\label{eq:int_Ramsey_ind}	
\end{align}
where the two Hamiltonians, $H_{\rm Ramsey}^{(\rm I)}(t)$ and $H_{\rm Ramsey}^{(\rm II)}(t)$, are defined such that Ramsey resonance occurs independently along each path, 
\begin{align}
	H_{\rm Ramsey}^{(\rm I)}(t):=
	\begin{cases}
	&-\frac{\omega}{2}\sigma_3 +\omega_1 \left(
	\cos\omega t \sigma_1 -\sin\omega t \sigma_2
	\right),~~~t_0\leq t < t_0+\tau/2,
	\\
	&-\frac{\omega_0}{2} \sigma_3,~~~~~~~~~~~~~~~~~~~~~~~~~~~~~~~~~~~~~t_0 +\tau/2\leq t <t_0 +\tau/2 +T,
	\\
	&-\frac{\omega}{2}\sigma_3 +\omega_1 \left(
	\cos\omega t \sigma_1 -\sin\omega t \sigma_2
	\right),~~~t_0+\tau/2+T\leq t < t_0+\tau+T,
	\end{cases}
\end{align}
and
\begin{align}
	H_{\rm Ramsey}^{(\rm II)}(t):=
	\begin{cases}
	&+\frac{\omega}{2}\sigma_3 +\omega_1 \left(
	\cos\omega t \sigma_1 +\sin\omega t \sigma_2
	\right),~~~t_0\leq t < t_0+\tau/2,
	\\
	&+\frac{\omega_0}{2} \sigma_3,~~~~~~~~~~~~~~~~~~~~~~~~~~~~~~~~~~~~~t_0 +\tau/2\leq t <t_0 +\tau/2 +T,
	\\
	&+\frac{\omega}{2}\sigma_3 +\omega_1 \left(
	\cos\omega t \sigma_1 +\sin\omega t \sigma_2
	\right),~~~t_0+\tau/2+T\leq t < t_0+\tau+T.
	\end{cases}
\end{align}
As in the case of the Rabi resonance, the total Hamiltonian $H_{\rm Ramsey}^{(\rm Ind)}(t)$ in Eq.~\eqref{eq:int_Ramsey_ind} becomes proportional to the identity operator ${|\rm I\rangle} {\langle \rm I|} + {|\rm II\rangle} {\langle \rm II|}= I_{\mathcal{P}}$ when we choose $H_{\rm Ramsey}^{(\rm I)}(t) = H_{\rm Ramsey}^{(\rm II)}(t)$, and in that event the system reduces to the direct measurement described by Eq.~\eqref{eq:tevco}.  In other words,  
the direct measurement can be regarded as a special case of the indirect weak measurement of this unified framework also in the Ramsey resonance.

From the above Hamiltonians, the time evolution operator becomes
\begin{align}
	e^{-i \int_{t_0}^{t_0+\tau+T} dt H_{\rm Ramsey}^{(\rm Ind)}(t)}
	&=U^{(\rm I)}(t_0,\,t_0+\tau+T)\otimes {| \rm I\rangle}{\langle \rm I|}+U^{(\rm II)}(t_0,\,t_0+\tau+T)\otimes {| \rm II\rangle}{\langle \rm II|},
\end{align}
where, for brevity, we defined two unitary operators as
\begin{align}
	U^{(\rm I)}(t_0,\,t_0+\tau+T):&= e^{i \omega (t_0 +\tau+T)\sigma_3/2}e^{-i \omega_1 \tau \sigma_1/2} e^{-i (\omega-\omega_0)T \sigma_3/2} e^{-i \omega_1 \tau \sigma_1/2}e^{-i \omega t_0 \sigma_3 /2},
	\\
	U^{(\rm II)}(t_0,\,t_0+\tau+T):&= e^{-i \omega (t_0 +\tau+T)\sigma_3/2}e^{-i \omega_1 \tau \sigma_1/2} e^{+i (\omega-\omega_0)T \sigma_3/2} e^{-i \omega_1 \tau \sigma_1/2}e^{+i \omega t_0 \sigma_3 /2}.
\end{align}
Let the initial state of the entire system be ${|\Lambda_i\rangle} = {|\psi_i \otimes \Psi_i\rangle}$, where ${|\psi_i\rangle}$ is the initial state of the spin and ${|\Psi_i\rangle}$ is the state of the path (probe).
The time evolution from $t_0$ to $t_0+\tau+T$ is
\begin{align}
	{|\Lambda_f\rangle}:=&\, e^{-i \int_{t_0}^{t_0+\tau+T} dt H_{\rm Ramsey}^{(\rm Ind)}(t)}{|\Lambda_i\rangle}\notag
 \\
 =&\,{\langle {\rm I}|\Psi_i\rangle} \left(U^{(\rm I)}(t_0,\, t_0+\tau+T)\otimes I_{\mathcal{P}}\right) {|\psi_i\otimes {\rm I}\rangle}
 +{\langle {\rm II}|\Psi_i\rangle} \left(U^{(\rm II)}(t_0,\, t_0+\tau+T)\otimes I_{\mathcal{P}}\right) {|\psi_i\otimes {\rm II}\rangle}.\label{eq:Ramsey_time_ind}
\end{align}

As in the Ramsey resonance under direct measurement, we choose ${|\psi_i\rangle}={|\pm\rangle}$, ${|\psi_f\rangle}={|\pm\rangle}$
and, 
after the time evolution given by Eq.~\eqref{eq:Ramsey_time_ind}, 
obtain the path state by implementing the post-selection,
\begin{align}
{|\Psi_f\rangle}&={\langle \psi_f|\Lambda_f\rangle}\notag
\\
&= {\langle \psi_f(\omega_1 \tau)|}{\rm exp} \bigg[-iT\frac{\omega-\omega_0}{2}\sigma_3\otimes P
\bigg] {|\psi_i(\omega_1\tau)\otimes \Psi_i (\omega (\tau+T))\rangle},\label{eq:Ramsey_time_ind2}
\end{align}
where ${|\psi_i(\omega_1\tau)\rangle}:=e^{-i\omega_1\tau \sigma_1/2}{|\psi_i\rangle}$, ${|\psi_f(\omega_1 \tau)\rangle}:=e^{i\omega_1 \tau \sigma_1/2}{|\psi_f\rangle}$ and
\begin{align}
{|\Psi_i \left(\omega (\tau+T)\right)\rangle}:=e^{\pm i\omega (\tau+T)/2} {\langle {\rm I}|\Psi_i\rangle}{|\rm I\rangle}+e^{\mp i\omega (\tau+T)/2}{\langle {\rm II}|\Psi_i\rangle}{|\rm II\rangle}, 
\end{align}
with $P$ given in Eq.~\eqref{eq:P}.
%
%
Regarding the time evolution operator in Eq.~\eqref{eq:Ramsey_time_ind2} as Eq.~\eqref{eq:Ham2}, we find
\begin{align}
    H_0=0,\qquad V=\frac{\phi_{\rm Ramsey}}{T}\sigma_3,
\end{align}
where $\phi_{\rm Ramsey}=(\omega-\omega_0) T/2$.
The coefficients in Eq.~\eqref{eq:h0V} are identified as
\begin{align}
    &h=0,\qquad n_0^{(h)}=0,\qquad \vec{n}^{(h)}=\left(0,\, 0,\, 1\right),
    \\
    &v=\phi_{\rm Ramsey}/T,\qquad n_0^{(v)}=0,\qquad \vec{n}^{(v)}=\left(0,\, 0,\, 1\right).
\end{align}
Combining these with Eq.~\eqref{eq:Psif_qubit}, we obtain
\begin{align}
    {|\Psi_f\rangle}\simeq {\langle \psi_f(\omega_1\tau)|\psi_i(\omega_1\tau)\rangle} \left[
    I_{\mathcal{P}}-ivT  \sigma^W_3(\omega_1) P
    \right] {|\Psi_i \left(\omega (\tau+T)\right)\rangle},
\end{align}
where
\begin{align}
    \sigma^W_3(\omega_1)=\frac{{\langle \psi_f(\omega_1\tau)|\sigma_3|\psi_i(\omega_1\tau)\rangle} }{{\langle \psi_f(\omega_1\tau)|\psi_i(\omega_1\tau)\rangle}}=\frac{{\langle \pm|e^{-i\omega_1 \tau\sigma_1/2}\sigma_3 e^{-i\omega_1 \tau\sigma_1/2}|\pm\rangle} }{{\langle \pm|e^{-i\omega_1 \tau \sigma_1}|\pm\rangle}}=\pm\frac{1}{\cos \left(\omega_1 \tau\right)}.\label{eq:ramset_ind_sigma_W}
\end{align}

From these results, the expectation values of the path observables, {\it e.g.,} $F=P_{\rm I}$ and $P_{\rm I+II}$ after the post-selection, are found to be
\begin{align}
    \Delta P_{\rm I}&={\langle P_{\rm I}\rangle}_f-{\langle P_{\rm I}\rangle}_i\simeq 2v T{\rm Im}\, \sigma^W_3(\omega_1) {\rm Cov}\left(P_{\rm I},\, P\right)_i,
    \\
    \Delta P_{\rm I+II}&={\langle P_{\rm I+II}\rangle}_f-{\langle P_{\rm I+II}\rangle}_i\simeq 
    vT\left[
    {\rm Re}\, \sigma^W_3(\omega_1){\langle C_{P_{\rm I+II} P}\rangle}_i
    +
    2{\rm Im}\, \sigma^W_3(\omega_1)
    {\rm Cov}\left(P_{\rm I+II} ,\,P\right)_i
    \right],\label{eq:del_P_I_II_ramsey}
\end{align}
where $vT=\phi_{\rm Ramsey}=(\omega-\omega_0)T/2$, and we used the notation,
\begin{align}
{\langle \mathcal{O}\rangle}_i
:=\frac{{\langle \Psi_i \left(\omega (\tau+T)\right)|\mathcal{O}|\Psi_i \left(\omega (\tau+T)\right)\rangle}}{{\langle \Psi_i \left(\omega (\tau+T)\right)|\Psi_i \left(\omega (\tau+T)\right)\rangle}},
\end{align}
for any operator $\mathcal{O}$.
Introducing a small parameter $\epsilon$ by $\omega_0 = \overline{\omega}_0 + \epsilon$, we find $vT = \epsilon T/2 = \delta_{\rm Ramsey}/2$ in the vicinity of the resonance point $\omega = \overline{\omega}_0$.
Here, $\delta_{\rm Ramsey}$ quantifies the measurement strength of the Ramsey resonance, as we have seen in the direct measurement in Section~\ref{sec:Ramsey_direct}. 
From these, one can retrieve the real and imaginary parts of the weak values,
\begin{align}
    {\rm Im}\, \sigma^W_3(\omega_1)&\simeq \frac{1}{2{\rm Cov}\left(P_{\rm I},\,P\right)_i}
    \frac{d \Delta P_{\rm I}}{d(vT)}\bigg|_{v=0},
    \\
    {\rm Re}\, \sigma^W_3(\omega_1)&\simeq
    \frac{1}{{\langle C_{P_{\rm I+II} P}\rangle}_i}\left(\frac{d \Delta P_{\rm I+II}}{d(vT)}\bigg|_{v=0}
    -
    \frac{{\rm Cov}\left(P_{\rm I+II},\,P\right)_i}{{\rm Cov}\left(P_{\rm I},\,P\right)_i}
    \frac{d \Delta P_{\rm I}}{d(vT)}\bigg|_{v=0}
    \right).
\end{align}
One then observes that, as in the case of the Rabi resonance, the weak value in Eq.~\eqref{eq:ramset_ind_sigma_W} also exhibits enhancement ({\it i.e.}, weak value amplification) in the limit $\omega_1 \tau = \pi/2$ which corresponds to the resonance peak.
This enhancement leads to a significant shift in the probe system, as described in Eq.~\eqref{eq:del_P_I_II_ramsey}, which can be attributed to the large weak value.
This shows that, as in the Rabi case, the indirect measurement allows us to obtain the real part of the weak value as well as the imaginary part, in contrast to the direct measurement scheme realized by putting $H_{\rm Ramsey}^{(\rm I)}(t) = H_{\rm Ramsey}^{(\rm II)}(t)$ in our unified framework where only the imaginary part arises.

Finally, we compare the Rabi and Ramsey resonances in the context of the indirect measurement scheme.
Since both of the resonances are sensitive only to the real part of the weak values, we may focus on $\Delta P_{\rm I+II}$ near the resonance point $\omega = \overline{\omega}_0$.
Near the maximal resonance peak ($\omega_1 t = \pi/2$ for the Rabi resonance and $\omega_1 \tau = \pi/2$ for the Ramsey resonance), Eqs.~\eqref{eq:rabi_delP_I_II} and \eqref{eq:del_P_I_II_ramsey} reduce to
\begin{align}
    \Delta P_{\rm I+II}^{\rm Rabi}\left(\epsilon\right)&\simeq \frac{\delta_{\rm Rabi}}{2}\,{\rm Re}\, \sigma^W_{a,L}\, {\langle C_{P_{\rm I+II}P}\rangle}_i,\quad \Delta P_{\rm I+II}^{\rm Ramsey}\left(\epsilon\right)\simeq
    \frac{\delta_{\rm Ramsey}}{2}\, {\rm Re}\,\sigma^W_3\, {\langle C_{P_{\rm I+II}P}\rangle}_i,
\end{align}
where $\delta_{\rm Rabi}=2\epsilon t/\pi$, $\delta_{\rm Ramsey}=\epsilon T$, and ${\rm Re}\,\sigma^W_{a,L}\simeq {\rm Re}\,\sigma^W_{3}$ near $\omega_1 t=\pi/2$ and $\omega_1\tau=\pi/2$.
We thus learn from the relation $\delta_{\rm Rabi}=2\delta_{\rm Ramsey}/\pi$ that, for $t=T$, even under the indirect measurement scheme the Ramsey resonance remains more sensitive to small deviations from the resonance point than the Rabi resonance does.

\section{Summary and Discussions}
\label{sec:sum}
In this paper, we revisited both the direct and indirect measurement schemes of weak measurement designed to gain the weak value of a physical observable and provided a unified framework for them.  
We first discussed that, in the direct weak measurement scheme, only the imaginary part of the weak value can be retrieved unless non-Hermitian time evolutions are introduced.
In contrast, the indirect measurement scheme, which considers an additional probe system and has 
been used to discuss the weak value amplification much more widely than the direct one, enables us to obtain both real and imaginary parts. 
Despite this difference, we have shown that the two schemes can be embedded within a single framework after reformulation of the two.

Armed with this, we analyzed spin-involved resonances to demonstrate that the amplifying phenomena observed in the Rabi resonance and the Ramsey resonance share the same character with those observed in the weak value amplification.  This results not only put our previous agreement~\cite{Ueda:2023xoj} in a firmer basis, but also clarify the underlying structure of these resonances as well as the conditions under which the resonances are enhanced.  
The correspondence between the direct and indirect schemes presented in this paper suggests that any resonance observed in one scheme has a counterpart in the other, opening the possibility of discovering new resonances beyond the direct measurement scheme known so far.

Looking ahead, we expect that our framework can offer a new perspective for precision measurements of fundamental parameters, such as the neutron EDM, couplings between dark matter (DM) and Standard Model particles, and the DM mass.  Indeed, 
while neutron EDM experiments using indirect weak measurement have already been studied~\cite{Ueda:2020qny}, approaches based on quantum resonances may offer improved robustness in spin preparation and to leverage established experimental techniques.
Also, revisiting these measurements with a focus on phase shifts near resonance peaks could enhance sensitivity.
Moreover, DM searches exploiting EDM-based measurements, in particular when axion-like particles couple to neutrons, may benefit from the indirect weak measurement techniques discussed here.

\section*{Acknowledgments}
DU would like to thank the CERN Theory Department for their financial support and hospitality during this work.
DU is supported by grants from the ISF (No.~1002/23 and 597/24) and the BSF (No.~2021800), and IT is supported by the JSPS KAKENHI grant (No.~25K07176).

\bibliography{Qres.bib}

\end{document}